\documentstyle[11pt,aaspp4]{article}

\input{psfig.sty}

\newcommand\sgra {SGR~1806$-$20}

\newcommand\sgrc {SGR~1900$+$14}

\newcommand\axpa {1E~2259$+$586}

\newcommand\rxte {{\it RXTE}}
\newcommand\asca {{\it ASCA}}
\newcommand\sax {{\it BeppoSAX}}
\newcommand\cxo {{\it Chandra}}
\newcommand\xmm {{\it XMM-Newton}}

\begin{document}

\title{The Prelude to and Aftermath of the Giant Flare of 2004 December 27:
Persistent and Pulsed X-ray Properties of SGR~1806$-$20 from 1993 to 2005}

\author{
Peter~M.~Woods\altaffilmark{1,2,3},
Chryssa~Kouveliotou\altaffilmark{3,4},
Mark~H.~Finger\altaffilmark{2,3},
Ersin~{G\"o\u{g}\"u\c{s}}\altaffilmark{5},
Colleen~A.~Wilson\altaffilmark{3,4},
Sandeep~K.~Patel\altaffilmark{2,3,4},
Kevin~Hurley\altaffilmark{6}, and
Jean~H.~Swank\altaffilmark{7}
}

\altaffiltext{1}{Dynetics, Inc. 1000 Explorer Blvd. Huntsville, AL 35806;
Peter.Woods@dynetics.com}
\altaffiltext{2}{Universities Space Research Association}
\altaffiltext{3}{National Space Science and Technology Center, 320 Sparkman Dr. 
Huntsville, AL 35805}
\altaffiltext{4}{NASA at Marshall Space Flight Center, VP62, Huntsville, AL
35812}
\altaffiltext{5}{Sabanci University, FENS, Istanbul 34956 Turkey}
\altaffiltext{6}{University of California at Berkeley, Space Sciences
Laboratory, 7 Gauss Way, Berkeley, CA 94720-7450}
\altaffiltext{7}{NASA at Goddard Space Flight Center, LHEA, Mail Code 662,
Greenbelt, MD 20771}

\begin{abstract}

On 2004 December 27, a highly-energetic giant flare was recorded from the
magnetar candidate \sgra.  In the months preceding this flare, the persistent
X-ray emission from this object began to undergo significant changes.  Here, we
report on the evolution of key spectral and temporal parameters prior to and
following this giant flare.  Using the {\it Rossi X-ray Timing Explorer}, we
track the pulse frequency of \sgra\ and find that the spin-down rate of this
SGR varied erratically in the months before and after the flare.  Contrary to
the giant flare in \sgrc, we find no evidence for a discrete jump in spin
frequency at the time of the December 27${\rm th}$ flare
($|\frac{\Delta\nu}{\nu}| < 5 \times 10^{-6}$).  In the months surrounding the
flare, we find a correlation between pulsed flux and torque consistent with the
model for magnetar magnetosphere electrodynamics proposed by Thompson, Lyutikov
\& Kulkarni (2002).  As with the flare in \sgrc, the pulse morphology of \sgra\
changes drastically following the flare.  Using the {\it Chandra X-ray
Observatory} and other publicly available imaging X-ray detector observations,
we construct a spectral history of \sgra\ from 1993 to 2005.  As we reported
earlier, the X-ray spectrum of \sgra\ cannot be fitted to a simple power-law
model.  The usual magnetar persistent emission spectral model of a power-law
plus a blackbody provides an excellent fit to the data.  We confirm the earlier
finding by Mereghetti et al.\ (2005a) of increasing spectral hardness of \sgra\
between 1993 and 2004.  We find both an increase in blackbody temperature and a
flattening of the power-law photon index.  However, our results indicate
significant differences in the temporal evolution of the spectral hardening. 
Rather than a direct correlation as proposed by Mereghetti et al., we find
evidence for a sudden torque change that preceded a gradual hardening of the
energy spectrum on a timescale of years.  Interestingly, the spectral hardness,
spin-down rate, phase-averaged, and pulsed flux of \sgra\ all peak months
before the flare epoch.

\end{abstract}



\section{Introduction}

Soft Gamma Repeaters (SGRs) are persistent, pulsed X-ray sources that
sporadically enter burst active episodes, or outbursts, lasting anywhere from a
few weeks to several months.  These outbursts in SGRs are composed of ordinary,
repetitive bursts and, in rare cases, flares.  The common bursts typically last
$\sim$0.1 s and reach peak luminosities up to $\sim$10$^{41}$ ergs s$^{-1}$,
while the flares have longer durations (up to $\sim$5 minutes) and generally
higher peak luminosities reaching $\sim$10$^{47}$ ergs s$^{-1}$.  From the
relatively dim persistent X-ray emission ($L_{\rm x}$ $\sim10^{33}-10^{35}$
ergs s$^{-1}$) to the brightest flares, the radiative output from SGRs spans
some 14 orders of magnitude making this class of objects the most energetically
dynamic among isolated neutron stars.  For a review of SGRs and Anomalous X-ray
Pulsars (AXPs), a class of objects closely related to SGRs, see Woods \&
Thompson (2006).

It is generally believed that SGRs and AXPs are magnetars (Thompson \& Duncan
1995, 1996), neutron stars with superstrong magnetic fields of order
$10^{14}-10^{15}$ G (Kouveliotou et al.\ 1998), whose bright X-ray emission is
powered by the decay of the strong field.  The persistent X-ray emission from
magnetars is believed to be due to magnetospheric currents driven by twists in
the evolving magnetic field (Thompson, Lyutikov \& Kulkarni 2002) and thermal
emission from the stellar surface (\"Ozel 2003; Ho \& Lai 2003; Zane et al.\
2001) heated by the decay of the strong field (Thompson \& Duncan 1996).  X-ray
pulsations arise from anisotropic emission from a stellar surface of presumably
non-uniform temperature in combination with strong gradients in the photon
opacity versus magnetic latitude (Thompson et al.\ 2002).  Recent detections of
hard X-ray emission (20$-$200 keV) from \sgra\ (Mereghetti et al.\ 2005b;
Molkov et al.\ 2005) show that the energy output is dominated by the
non-thermal (magnetospheric) component.  Their burst emission results from
either a build up of magnetic stress and eventual release of this energy
through fracturing of the crust (Thompson \& Duncan 1995) or by magnetic
reconnection within the stellar magnetosphere (Lyutikov 2003).  In both burst
trigger schemes, the result is a trapped pair-photon fireball which cools and
radiates giving rise to the burst.

Burst active episodes in \sgrc, in particular outbursts containing flares, have
shown a measureable impact on the spectral and temporal properties of the
underlying persistent X-ray source.  For example, \sgrc\ entered a phase of
intense burst activity in 1998 May that included a giant flare recorded on 1998
August 27 (Hurley et al.\ 1999; Feroci et al.\ 2001).  Early in this outburst
(May $-$ June), the pulsed flux from the SGR was enhanced by a factor $\sim$2
above its nominal pre-outburst level (Woods et al.\ 2001).  Unfortunately,
there was a three month gap in pointed X-ray observations of the source prior
to the giant flare, so very little is known about the pre-flare flux
evolution.  During and following the flare, there was a sudden rise in the soft
X-ray persistent/pulsed flux from the SGR and a dramatic change in pulse shape
(Woods et al.\ 2001).  The flux increase, or X-ray afterglow, decayed rapidly
as a power-law in time over the next $\sim$40 days and has been attributed to
the heating of the outer crust of a neutron star with a 10$^{15}$ G surface
field (Lyubarsky, Eichler \& Thompson 2002).  The pulse profile change,
however, has persisted for at least three years following the flare, likely
indicative of a sustained rearrangement of the external field geometry (Woods
et al.\ 2001; {G\"o\u{g}\"u\c{s}} et al.\ 2002).  Further instances of flux
enhancements and spectral variability in this SGR have been observed following
less-energetic intermediate flares (Ibrahim et al.\ 2001; Feroci et al.\ 2003;
Lenters et al.\ 2003).  The interplay between burst activity in \sgrc\ and the
persistent emission properties has provided useful insight into its nature and
by association, the nature of magnetars in general.

Starting in 2004 May, \sgra\ entered a phase of enhanced burst activity that
has persisted for at least one year.  Over the course of this outburst, more
than 300 bursts were recorded from all-sky instruments within the
Interplanetary Network (IPN).  The pinnacle of this burst active episode was a
giant flare recorded on 2004 December 27 (Hurley et al.\ 2005; Palmer et al.\
2005; Mereghetti et al.\ 2005c), the brightest gamma-ray transient ever
observed, briefly brighter than any observed solar flare.  This giant flare had
a peak luminosity of $\sim$2 $\times$ 10$^{47}$ ergs s$^{-1}$, a total energy
of $\sim$5 $\times$ 10$^{46}$ ergs, and a duration of $\sim$5 minutes. 
Following this flare was a long-lived radio afterglow caused by the outflow of
material from the star during the flare (Gaensler et al.\ 2005; Cameron et al.\
2005; Gelfand et al.\ 2005; Taylor et al.\ 2005; Fender et al.\ 2006).

Here, we present a comprehensive spectral and temporal history of the
persistent X-ray emission from \sgra\ leading up to and following the giant
flare.  We discuss correlations between variability in the persistent X-ray
source and burst activity and the implications these have for the burst/flare
trigger.  Specifically, we report on X-ray observation of \sgra\ performed with
the {\it Rossi X-ray Timing Explorer (RXTE)} and the {\it Chandra X-ray
Observatory (Chandra)} between 2001 January and 2005 April.  From these data,
we extend the pulse frequency and morphology history of the source 4$-$5 years
beyond our earlier work (Woods et al.\ 2002; {G\"o\u{g}\"u\c{s}} et al.\ 2002),
and by inclusion of archival data, construct a spectral history of \sgra\
between 1993 and 2005.

\section{Observations}

We have observed \sgra\ 194 separate occasions with \rxte\ between 2001 January
1 and 2005 April 11 as part of our ongoing monitoring and Target-of-Opportunity
(ToO) campaigns.  A complete list of \rxte\ observations can be retrieved from
the archive maintained by the High Energy Astrophysics Science Archive Research
Center\footnote{http://heasarc.gsfc.nasa.gov}.  The sampling of the \rxte\
observations depended primarily upon the behavior of the source.  During
intense burst active episodes or when the persistent source was relatively
bright, the sampling was much higher.  For example, during a 6-month interval
between 2004 May and November prior to the giant flare when the source was very
active, \rxte\ observed \sgra\ 85 times.  The time intervals covered by these
observations can be found in Tables~\ref{tab:xtespin}~and~\ref{tab:spline}.

The configuration of the PCA and the High-Energy X-ray Timing Experiment
(HEXTE) instruments was optimized to study both the persistent (pulsed)
emission and burst emission from the SGR.  For the PCA instrument, the data
used in the analysis of the persistent emission described here were acquired in
event mode {\tt E\_125us\_64M\_0\_1s} prior to 2004 June 29 and in {\tt
GoodXenon\_2s} mode thereafter.  There are a handful of exceptions to this rule
caused ordinarily by rapid response to ToO triggers and the inability to change
data modes on a very short timescale.  For the High-Energy X-ray Timing
Experiment (HEXTE) instrument, the data used here were acquired in either {\tt
E\_8us\_256\_DX1F} or {\tt E\_8us\_256\_DX0F} mode, ordinarily without rocking
the HEXTE clusters (i.e.\ staring mode).

We have observed \sgra\ five times with \cxo\ between 2003 July and 2004
October as part of our ToO program.  One additional observation of \sgra\ with
\cxo\ was carried out on 2005 February 8 following the giant flare (Rea et al.\
2005).  We will include an independent analysis of this data set here for
completeness.  In each of these observations, the Advanced CCD Imaging
Spectrometer (ACIS) was used as the focal-plane detector.  The SGR was
positioned on the S3 chip at the nominal aimpoint.  The ACIS chips were
operated in Continuous Clocking (CC) mode which sacrifices one dimension of
spatial resolution for improved time resolution of 2.85 ms.  The CC-mode was
employed in order to avoid pulse pile-up and allow study of the pulsations and
bursts.  Details of these observations are presented in Table~\ref{tab:cxoobs}.


\begin{table}[!h]
\begin{minipage}{1.0\textwidth}
\begin{center}
\caption{\cxo\ observation log for \sgra\ between 2003 July and 2005 February. 
\label{tab:cxoobs}} 
\vspace{10pt}
\begin{tabular}{cccc} \hline \hline

 Name & \cxo\ Sequence &     Date     & Source Exposure  \\
      &    Number      &              &   (ks)    \\\hline
 
 CXO1 &    500412      & 2003 Jul 03  &   25.1    \\
 CXO2 &    500464      & 2004 May 27  &   50.2    \\
 CXO3 &    500465      & 2004 Jun 22  &   20.2    \\
 CXO4 &    500462      & 2004 Aug 13  &   35.2    \\
 CXO5 &    500463      & 2004 Oct 09  &   35.2    \\
 CXO6 &    500597      & 2005 Feb 08  &   29.1    \\

\hline\hline
\end{tabular}
\end{center}
\end{minipage}\hfill
\end{table}

The \rxte\ PCA observations allow us to precisely track the pulse frequency,
pulse morphology, and pulsed flux of the persistent X-ray emission while the
\cxo\ observations measure the spectral parameters and pulsed fraction.  In
this sense, the two data sets are quite complementary providing a comprehensive
picture of the state of the source at each common epoch.

Several hundred bursts were recorded from \sgra\ within the \rxte\ data
presented here and $\sim$40 bursts were detected during the \cxo\
observations.  Many of the bursts detected with \cxo\ were also recorded with
\rxte\ which enables us to perform joint spectral analysis.  Scientific results
obtained using the burst data detected during these observations will be
presented in subsequent papers (e.g.\ {G\"o\u{g}\"u\c{s}} et al.\ in
preparation). The bursts have been removed from all data analysis reported
here.

\section{Pulse Timing}

Previously, we compiled a pulse frequency and frequency derivative history of
\sgra\ between 1993 and 2000 (Woods et al.\ 2002).  We found that the spin-down
rate of this SGR was relatively stable between 1993 and 2000 January.  During
the first half of 2000, the spin-down rate increased by a factor $\sim$4, a
large and sudden jump that persisted through at least the beginning of 2001. 
Precision timing of the SGR before and after the large change in spin down
revealed strong timing noise on a wide array of time scales (Woods et al.\
2000, 2002).  In this section, we report on frequency and frequency derivative
measurements from 2001 to present using the \rxte\ PCA and \cxo\ ACIS data, and
thus extend our knowledge of the spin ephemeris of this SGR up through 2005
October.

In the analysis summarized below, we have followed techniques for measuring the
pulse frequency described in detail within earlier works (e.g.\ Woods et al.\
2002).  In general, we use an epoch-folding technique to measure the pulse
frequency and higher derivatives.  In this method, the data are split into
discrete intervals and folded on some trial frequency.  The resulting pulse
profile is cross-correlated with a high signal-to-noise template profile
(derived from long integrations of contemporaneous data) and a phase offset is
measured.  The phase offsets from each interval are fit to either a low-order
polynomial or a quadratic spline, depending upon the data set.  The fits to the
measured phase offsets yield the spin ephemeris for the SGR within the
specified time range.  A new template profile is constructed using this
ephemeris and the procedure is iterated until the fit parameters converge. 
This procedure ordinarily only requires one iteration.

\subsection{\cxo\ Timing}

For each of the six \cxo\ ACIS observations of \sgra, we started with the
standard level 2 filtered event list.  First, we found the centroid for the
peak of the one-dimensional image from each Continuous Clocking (CC) mode
observation and selected counts within 4 pixels of the centroid.  We further
selected counts with measured energies between 0.5 and 7.0 keV and constructed
a light curve with 0.5 s resolution.  Bursts were identified as bins having a
number of counts such that the normalized Poisson probability of chance
occurrence was less than 1\%.  In cases where we had simultaneous coverage with
\rxte, the bursts were first confirmed within the PCA light curve.  We
identified and removed a total of $\sim$40 bursts from the six \cxo\
observations of \sgra.

Once the data were cleaned, we corrected the CC-mode time tags to the true
photon arrival time\footnote{http://wwwastro.msfc.nasa.gov/xray/ACIS/cctime/}
and barycenter corrected these times using {\tt axbary}.  Next, we searched for
the pulse frequency using the $Z^2_2$ statistic.  The pulse frequency of \sgra\
showed up clearly in all observations except during the observation directly
following the giant flare (CXO6).  During that observation, the pulsed fraction
was extremely small, making the pulsed signal undetectable (Rea et al.\ 2005). 
Using the epoch-folding technique described above, we refined our pulse
frequency measurement for each observation.  The pulse frequencies are listed
in Table~\ref{tab:spin}.


\begin{table}[!h]
\begin{minipage}{1.0\textwidth}
\begin{center}
\caption{Pulse ephemerides for \sgra\ derived from \cxo\
observations between 2003 August and 2004 October.  Numbers given in 
parentheses indicate the 1$\sigma$ error in  the least significant digit(s). 
\label{tab:spin}} 
\vspace{10pt}
\begin{tabular}{cccl} \hline \hline

 Observation &   Epoch   &       Time Range        &    $\nu$      \\
   Label     & (MJD TDB) &       (MJD TDB)         &     (Hz)       \\\hline
 
 CXO1       & 52854.658 & 52854.514 $-$ 52854.806 & 0.1326803(15) \\
 CXO2       & 53152.896 & 53152.606 $-$ 53153.193 & 0.1324527(5) \\
 CXO3       & 53178.718 & 53178.605 $-$ 53178.832 & 0.132423(4) \\
 CXO4       & 53230.460 & 53230.266 $-$ 53230.664 & 0.1323718(6) \\
 CXO5       & 53287.220 & 53287.017 $-$ 53287.416 & 0.1323219(12) \\

\hline\hline
\end{tabular}
\end{center}
\end{minipage}\hfill
\end{table}

\subsection{\rxte\ Timing}

The \rxte\ PCA data were first screened to remove bursts and instrumental
background flares seen within individual PCUs that ordinarily occur when the
high voltage is being switched on or off.  The screened event lists were
filtered on energy (2$-$10 keV) and barycenter corrected using {\tt faxbary}.

Since 2001 January, there have been more than one hundred pointed observations
of \sgra\ with \rxte, most of which occurred during the 2004$-$2005 burst
active episode.  These observations were carefully scheduled to allow for phase
connection across intervals of weeks to months.  Due to the strong timing noise
in this SGR, the gaps between pointings within a given observing campaign could
not exceed $\sim$1 week.

We have grouped these observations into 18 separate intervals.  For each
interval, the data were grouped into segments long enough to accurately measure
the pulse phase.  The exposure times for these segments were 3$-$10 ks,
depending upon the pulsed amplitude of the SGR at the time.  With the exception
of the longest interval in 2004, we were able to fit the segment pulse phases
to low-order polynomials.  The parameters for the 17 polynomial fits are listed
in Table~\ref{tab:xtespin}.


\begin{table}[!h]
\begin{minipage}{1.0\textwidth}
\begin{center}
\caption{Pulse frequency ephemerides for \sgra\ derived from \rxte\ PCA
observations between 2001 January and 2005 August.  Numbers given in 
parentheses indicate the 1$\sigma$ error in  the least significant digit(s). 
\label{tab:xtespin}} 
\vspace{10pt}
\begin{tabular}{cclll} \hline \hline

   Epoch$^{a}$   &  Time Range & $\nu$ & $\dot{\nu}$ & $\ddot{\nu}$  \\
 (MJD TDB) &  (MJD TDB)  &  (Hz) & ($10^{-12}$ Hz s$^{-1}$) & ($10^{-18}$ Hz s$^{-2}$) \\

\hline
 
 52022.549 & 52021.560 $-$ 52023.501 & 0.1333027(4)   &  ...     & ...    \\ 
 52098.000 & 52092.810 $-$ 52102.634 & 0.13324616(13) & -8.9(8)  & ...    \\ 
 52224.000 & 52215.051 $-$ 52236.003 & 0.13315465(4)  & -8.92(9) & ...    \\ 
 52302.069 & 52301.818 $-$ 52302.270 & 0.133093(2)    &  ...     & ...    \\ 
 52559.344 & 52559.269 $-$ 52559.419 & 0.132900(11)   &  ...     & ...    \\ 
 52854.663 & 52854.553 $-$ 52854.770 & 0.132682(4)    &  ...     & ...    \\ 
 52871.963 & 52871.368 $-$ 52872.529 & 0.1326739(7)   &  ...     & ...    \\ 
 52893.273 & 52893.197 $-$ 52893.349 & 0.132654(4)    &  ...     & ...    \\ 
 52927.799 & 52926.352 $-$ 52929.263 & 0.1326232(2)   &  ...     & ...    \\ 
 53027.062 & 53026.886 $-$ 53027.242 & 0.132560(3)    &  ...     & ...    \\ 
 53051.500 & 53050.666 $-$ 53057.050 & 0.1325295(2)   & -9.0(8)  & ...    \\ 
 53078.938 & 53078.885 $-$ 53078.990 & 0.132510(8)    &  ...     & ...    \\ 
 53097.477 & 53096.424 $-$ 53098.023 & 0.1324902(3)   &  ...     & ...    \\ 
 53395.000 & 53392.865 $-$ 53410.202 & 0.13227473(3)  & -3.23(5) & ...    \\ 
 53435.915 & 53435.707 $-$ 53436.265 & 0.1322633(7)   &  ...     & ...    \\ 
 53460.000 & 53450.829 $-$ 53470.901 & 0.13225498(1)  & -2.86(3) & ...    \\ 
 53555.000 & 53545.565 $-$ 53565.586 & 0.13222717(4)  & -4.73(5) & -1.3(4) \\ 
 
\hline\hline
\end{tabular}
\end{center}
\begin{small}
\noindent$^{a}$ Many observations were either too short or the frequency change
too small to allow us to measure $\dot{\nu}$ and/or $\ddot{\nu}$.  In these
instances, the corresponding table entries were left blank (i.e.\ ``...''). \\
\end{small}
\end{minipage}\hfill
\end{table}

We observed \sgra\ 68 times across a 182 day interval between 2004 May 24 and
2004 November 22 with an average and maximum separation of 2.7 and 7.9 days
between consecutive pointings, respectively.  We were able to phase connect
portions of this interval lasting up to $\sim$2 months using our standard
approach involving polynomial fitting and extrapolating the polynomial to the
epoch of the next observation usually a few days later.  However, as the degree
of the polynomial increased beyond 4$^{\rm th}$ order, the extrapolation became
problematic and we could no longer identify the correct number of cycle counts
to the next epoch.  This led us to develop a new approach to phase connect long
stretches of ``noisy'' pulsar data.

We developed a least-squares fitting routine that uses the measured phases and
frequencies at each of the 68 epochs and fits for the optimal cycle slips
between the epochs, and in turn, yields a cubic spline solution to the full
span covered by the data (see Appendix for details).  We first measured phases
at each epoch assuming an average frequency and frequency derivative for the
full interval.  Next, we measured frequencies at these epochs by splitting
individual segments into three sections of equal exposure (1$-$3 ks), measuring
phases for each section, and fitting the phases to a line.  Rather than fitting
the full 182 day interval at once to a high-order polynomial, we chose to fit
smaller time spans (30 days) to a quadratic and step the 30 day window through
the time interval in steps of 15 days.  Each 30 day window typically contained
10 observing epochs.  Within each window, we compared the measured $\chi^2$ of
the best fit to the next best solution.  The change in $\chi^2$ between the two
solutions ranged between 31 and 760 with an average $\Delta \chi^2$ of 245. 
The average number of degrees of freedom for each fit was 7, thus we are
confident that we identified the proper cycle counts between most, probably all
epochs.  Once the absolute phases were determined, we fit the data to a
quadratic spline model of 26 segments of 7 days each.  Segments longer than
$\sim$10 days clearly required a cubic phase term to adequately model the
measured phases and achieve a reduced $\chi^2$ of $\sim$1.  The quadratic
spline fit parameters are listed in Table~\ref{tab:spline}.


\begin{table}[!p]
\begin{minipage}{1.0\textwidth}
\begin{center}
\caption{Pulse frequency ephemerides for \sgra\ derived from a quadratic spline 
fit to \rxte\ PCA observations between 2004 May and November and 2000 March
and June.  Numbers given in 
parentheses indicate the 1$\sigma$ error in the least significant digit. 
\label{tab:spline}} 
\vspace{10pt}
\begin{tabular}{cccc} \hline \hline

   Epoch   &       Time Range        &     $\nu$     & $\dot\nu$ \\
 (MJD TDB) &       (MJD TDB)         &     (Hz)      & (10$^{-12}$ Hz s$^{-1}$) \\\hline

 53153.019 & 53149.519 $-$ 53156.519 & 0.1324520(2)  &  -9.8(7)  \\
 53160.019 & 53156.519 $-$ 53163.519 & 0.13244590(3) & -10.3(2)  \\
 53167.019 & 53163.519 $-$ 53170.519 & 0.13243955(2) & -10.7(2)  \\
 53174.019 & 53170.519 $-$ 53177.519 & 0.13243230(2) & -13.2(2)  \\
 53181.019 & 53177.519 $-$ 53184.519 & 0.13242491(2) & -11.2(2)  \\
 53188.019 & 53184.519 $-$ 53191.519 & 0.13241799(2) & -11.7(2)  \\
 53195.019 & 53191.519 $-$ 53198.519 & 0.13241071(2) & -12.4(2)  \\
 53202.019 & 53198.519 $-$ 53205.519 & 0.13240326(3) & -12.2(2)  \\
 53209.019 & 53205.519 $-$ 53212.519 & 0.13239526(3) & -14.2(3)  \\
 53216.019 & 53212.519 $-$ 53219.519 & 0.13238710(3) & -12.7(3)  \\
 53223.019 & 53219.519 $-$ 53226.519 & 0.13237939(3) & -12.8(2)  \\
 53230.019 & 53226.519 $-$ 53233.519 & 0.13237121(3) & -14.2(2)  \\
 53237.019 & 53233.519 $-$ 53240.519 & 0.13236299(2) & -13.0(2)  \\
 53244.019 & 53240.519 $-$ 53247.519 & 0.13235586(3) & -10.6(2)  \\
 53251.019 & 53247.519 $-$ 53254.519 & 0.13234940(2) & -10.7(2)  \\
 53258.019 & 53254.519 $-$ 53261.519 & 0.13234307(2) & -10.2(2)  \\
 53265.019 & 53261.519 $-$ 53268.519 & 0.13233725(2) &  -9.1(2)  \\
 53272.019 & 53268.519 $-$ 53275.519 & 0.13233188(3) &  -8.7(3)  \\
 53279.019 & 53275.519 $-$ 53282.519 & 0.13232660(3) &  -8.8(3)  \\
 53286.019 & 53282.519 $-$ 53289.519 & 0.13232167(3) &  -7.6(2)  \\
 53293.019 & 53289.519 $-$ 53296.519 & 0.13231746(2) &  -6.3(2)  \\
 53300.019 & 53296.519 $-$ 53303.519 & 0.13231369(3) &  -6.1(2)  \\
 53307.019 & 53303.519 $-$ 53310.519 & 0.13230984(3) &  -6.6(3)  \\
 53314.019 & 53310.519 $-$ 53317.519 & 0.13230610(3) &  -5.8(3)  \\
 53321.019 & 53317.519 $-$ 53324.519 & 0.13230256(6) &  -5.8(4)  \\
 53327.917 & 53324.519 $-$ 53331.315 & 0.1322994(2)  &  -5(1)    \\
\hline

 51626.778 & 51616.778 $-$ 51636.778 & 0.13355154(3) &  -6.02(7)  \\
 51646.778 & 51636.778 $-$ 51656.778 & 0.13354098(2) &  -6.20(5)  \\
 51666.778 & 51656.778 $-$ 51676.778 & 0.13352986(2) &  -6.67(5)  \\
 51686.778 & 51676.778 $-$ 51696.778 & 0.13351850(2) &  -6.49(6)  \\
 51708.332 & 51696.778 $-$ 51719.887 & 0.13350636(6) &  -6.5(2)  \\

\hline\hline
\end{tabular}
\end{center}
\end{minipage}\hfill
\end{table}

We decided to apply this technique to an archival \sgra\ data set from 2000
that we previously could not phase connect in its entirety.  In Woods et al.\
(2002), we reported two high-order spin ephemerides (2000a and 2000b in
Table~4) that covered portions of the 2000 \rxte\ data set.  Using this new
technique, we were successful in phase connecting more \rxte\ observations and
effectively extending the 2000a spin ephemeris to earlier times.  The new spin
ephemeris that now supercedes the 2000a spin ephemeris in Woods et al.\ (2002)
is given in Table~\ref{tab:spline}.

\subsection{Pulse Frequency History}

\begin{figure}[!p]

\centerline{
\psfig{file=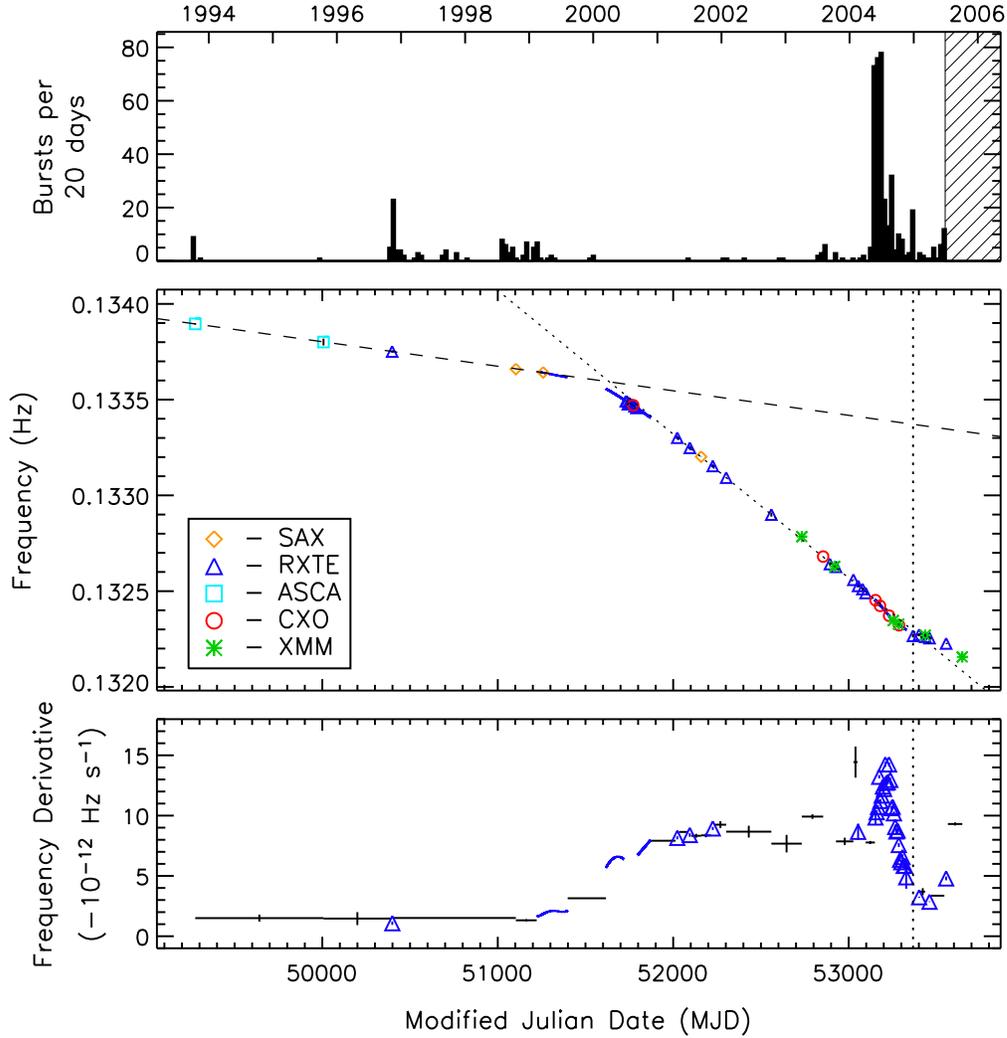,height=6.0in}}

\caption{Pulse frequency and frequency derivative history of \sgra\ between
1993 and 2005.  {\it Top} -- Burst rate history (through 2004 October) as seen
with instruments within the Inter-Planetary Network.  The time of the giant
flare is indicated in subsequent panels by a vertical black dotted line.  The
burst rate data are complete through 2005 June.  {\it Middle} -- Pulse
frequency history of the SGR as measured using an array of X-ray detectors (see
insert legend).  The dashed black line indicates a fit to frequency
measurements between 1993 and 2000 January ($\dot{\nu}$ = $-1.48 \times
10^{-12}$ Hz s$^{-1}$).  The diagonal dotted black line indicates a fit to
frequency measurements between 2001 January and 2004 April ($\dot{\nu}$ =
$-8.69 \times 10^{-12}$ Hz s$^{-1}$).  {\it Bottom} -- Pulse frequency
derivative history of the SGR.  Blue triangles indicate instantaneous frequency
derivative measurements made with the \rxte\ PCA.  Solid blue lines indicate
continuous frequency derivative measurements from high-order ($>$3) polynomial
fits to long stretches of phase-connected PCA observations.  Black lines
indicate average frequency derivative values between widely spaced frequency
measurements.  See text for details. \label{fig:longspin}}

\end{figure}

We constructed a comprehensive pulse frequency and frequency derivative history
of \sgra\ from 1993 to 2005 (Figure~\ref{fig:longspin}) by combining our
current \rxte\ and \cxo\ measurements with our earlier work (Woods et al.\
2002) and recently reported pulse frequency measurements derived from \xmm\
observations of the SGR (Mereghetti et al.\ 2005a; Tiengo et al.\ 2005; Rea et
al.\ 2005b).  For comparison, we included a histogram of the bursts recorded
with the Inter-Planetary Network (IPN) from \sgra\ in the top panel of
Figure~\ref{fig:longspin}.  Note that the bursts are of varying peak flux and
total fluence.  In general, the burst energies follow a power-law distribution
(e.g.\ {G\"o\u{g}\"u\c{s}} et al.\ 2000).  Although the detectors that make up
the IPN (and hence the IPN burst sensitivity) have changed over the last 12
years, we consider the IPN burst rate as a good indicator of overall burst
activity of the source.

In the period 1990$-$2005, 19 spacecraft contributed one or more instruments to
the IPN (PVO, Ginga, GRANAT, DMSP, Ulysses, GRO, Yohkoh, Eureca-A, Mars
Observer, Coronas, SROSS-C, Wind, HETE, BeppoSAX, NEAR, Mars Odyssey, RHESSI,
INTEGRAL, and Swift).  Between 5 and 11 of them were operating simultaneously,
depending on the exact date.  They had a wide variety of operating modes,
energy ranges, time resolutions, duty cycles, and planet-blocking constraints
for observing bursts from \sgra.  Some were capable of independently localizing
bursts, while others were not; bursts detected by the non-localizing
instruments could be traced to \sgra\ by triangulation, if they were observed
by at least two spacecraft.  An imaging instrument such as INTEGRAL-IBIS can
detect bursts with fluences as small as 7 $\times$ 10$^{-9}$ erg cm$^{-2}$
(Gotz et al.\ 2006), while one such as GRO-BATSE has a slightly higher
threshold ($\sim$1.4 x 10$^{-8}$ erg cm$^{-2}$ - [{G\"o\u{g}\"u\c{s}} et al.\
2000]).  When a two-spacecraft triangulation is required, the threshold
increases to several times 10$^{-7}$ erg cm$^{-2}$.  Because so many spacecraft
were operating simultaneously, this is a good approximation to the largest
fluence threshold for the 1990$-$2005 period.



Over the last 12 years, \sgra\ has undergone two epochs of relatively steady
spin down, but at very different rates.  Between 1993 and 2000 January, the
average spin-down rate was $-1.48 \times 10^{-12}$ Hz s$^{-1}$, or $\sim$6
times smaller than between 2001 January and 2004 April ($-8.69 \times 10^{-12}$
Hz s$^{-1}$).  The dramatic change in spin-down rate that began in 1999 and
lasted $\sim$2 years, occurred without any spectacular increase in burst
activity, change in persistent flux, pulse profile profile change, etc..

\begin{figure}[!p]

\centerline{
\psfig{file=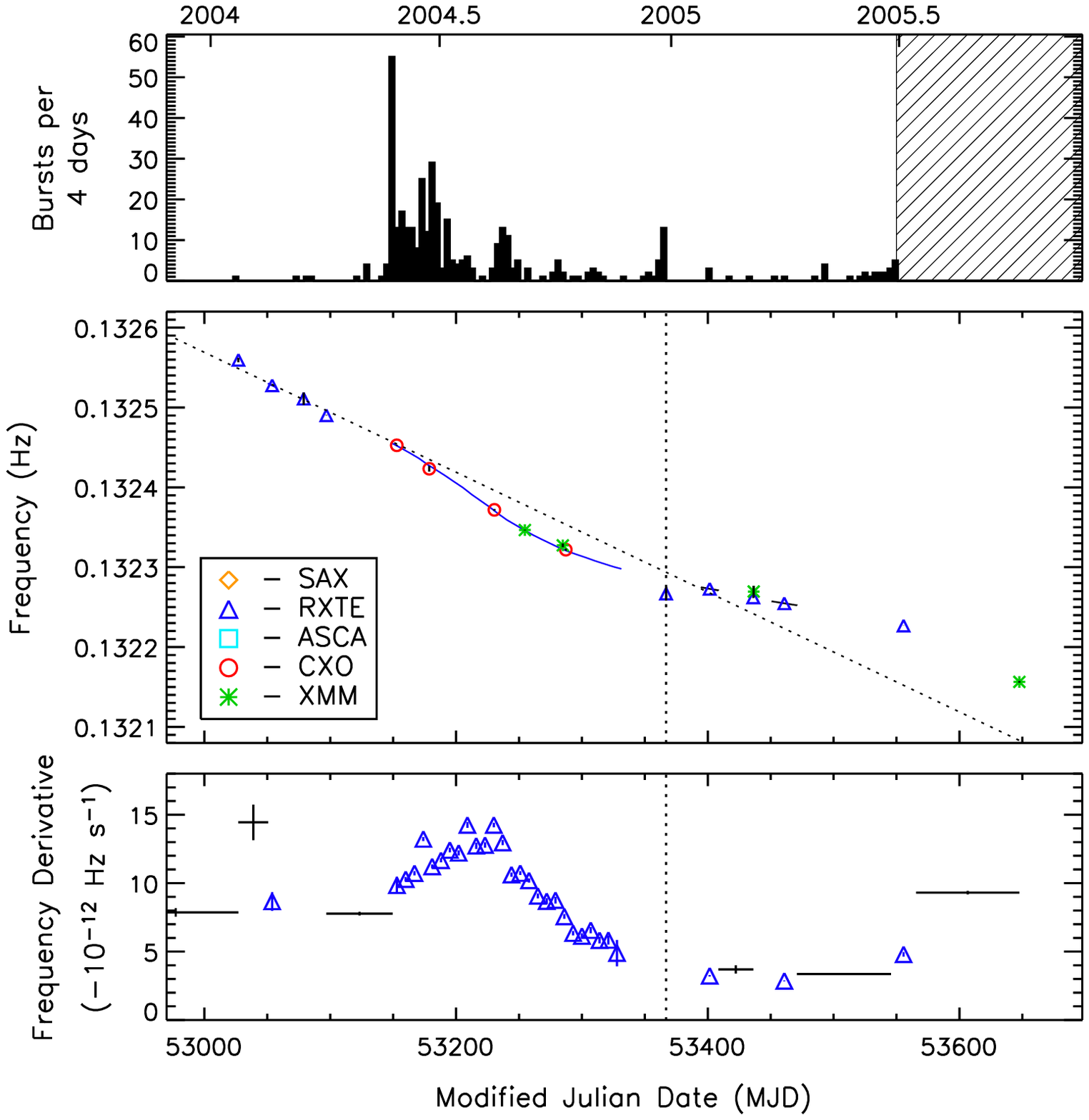,height=6.0in}}

\caption{Pulse frequency and frequency derivative history of \sgra\ during 2004
and 2005.  {\it Top} -- Burst rate history (through 2004 October) as seen with
instruments within the Inter-Planetary Network.  The time of the giant flare is
indicated in subsequent panels by a vertical dotted line.  The burst rate data
are complete through 2005 June.  {\it Middle} -- Pulse frequency history of the
SGR as measured using an array of X-ray detectors (see insert legend).  The
dotted black line indicates a fit to frequency measurements between 2001
January and 2004 April ($\dot{\nu}$ = $-8.69 \times 10^{-12}$ Hz s$^{-1}$). 
The solid blue line indicates the frequency evolution measured by the quadratic
spline fit to the \rxte\ observations during this interval.  {\it Bottom} --
Pulse frequency derivative history of the SGR.  Blue triangles indicate
instantaneous frequency derivative measurements made with the \rxte\ PCA. 
Black lines indicate average values between widely spaced frequency
measurements.  See text for details. \label{fig:shortspin}}

\end{figure}

Only during the months leading up to the giant flare did we begin to observe
large-amplitude, short-lived deviations from steady spin down
(Figure~\ref{fig:shortspin}).  \footnote{We note that pulse profile changes
were observed during this epoch (see \S4) and such changes can, in general,
influence the pulse timing solution. However, the pulse morphology changes were
small in the 2-10 keV energy band over which the pulse timing analysis was
carried out and the phase drifts would have had to have been extremely large
(of order multiple cycles per month) in order to account for the variability in
the frequency derivative.}  However, the frequency measurements between 2001
January and 2004 April were too sparse to detect similar frequency derivative
changes.  The spin-down rate of \sgra\ steadily dropped between 2004 August and
November.  After 2004 November 22, \rxte\ observations were suspended due to
Sun-angle constraints.  Note that the spin-down rate began dropping well before
the giant flare on 2004 December 27 (MJD 53366).  When we fit the frequency
derivative measurements between MJD 53150 and 53300 to a quadratic, we measure
a centroid of MJD 53209$\pm$1.  Thus, the torque on the star reached a maximum
$\sim$5 months prior to the giant flare.



There was no measurable discrete jump in frequency of either sign at the time
of the flare.  Extrapolating the last pre-flare and first post-flare
ephemerides to the time of the flare, we find an insignificant difference
between the two predicted frequencies of 3.1$\pm$2.0 $\times$ 10$^{-7}$ Hz
where the forward extrapolation yielded the larger expected frequency.  The
error reported here reflects the statistical error only and not the (dominant)
systematic error caused by the strong timing noise of \sgra.  Both
extrapolations are consistent with the relatively imprecise pulse frequency
measured during the tail of the flare itself (Woods et al.\ 2005).  During the
tail of the flare, the pulse profile changed dramatically and significantly
biased the pulse frequency measurement.  Thus, the formal 3$\sigma$ upper limit
on the size of a hypothetical flare-induced frequency jump is
$|\frac{\Delta\nu}{\nu}| < 5 \times 10^{-6}$.  This limit is more than one
order of magnitude smaller than the frequency jump inferred for \sgrc\ 
($\frac{\Delta\nu}{\nu} \approx -1 \times 10^{-4}$) at the time of the August
27 flare (Woods et al.\ 2001).  We caution that the necessary frequency
extrapolations employed here are susceptible to significant errors if the spin
down rate of \sgra\ changed significantly during the 63 day gap in
observations.  Moreover, this particular SGR has been known for some time to
exhibit strong timing noise (Woods et al.\ 2000).  However, the spin down
changes would have had to have been large in amplitude, short-lived in
duration, and precisely constructed in order to counter-balance a frequency
jump as large as $|\frac{\Delta\nu}{\nu}| \approx 1 \times 10^{-4}$ and still
give the appearance of no flare-induced frequency jump when viewed with the
existing data.  We consider this scenario highly improbable.

The pre-flare reduction in torque continued following the giant flare,
gradually approaching the pre-2000 spin-down rate 4$-$6 months following the
flare.  However, this trend quickly reversed itself one year after the flare
and the most recent spin-down rate is equal to the nominal rate seen between
2001 and 2004.

\section{Pulse Morphology Changes}

\subsection{Temporal Evolution}

{G\"o\u{g}\"u\c{s}} et al.\ (2002) investigated the pulse profile evolution of
\sgra\ between 1996 and 2001 using \rxte\ monitoring data.  During the first
couple weeks of the 1996 outburst, the 2$-$10 keV pulse profile of \sgra\
consisted of a broad, double-peaked pulse.  Due to Sun-angle pointing
constraints for \rxte, the source was not observed before the end of the
outburst.  At some point between 1996 November and 1999 February, the next time
this SGR was observed, the pulse profile of the SGR simplified to a single,
narrow pulse.  We note that the majority of the burst energy emitted during the
1996 outburst {\it followed} the sequence of PCA observations used to construct
this pulse profile.  Thus, it is not known whether the pulse shape change
happened suddenly during the intense portions of the 1996 outburst, or if the
change was more gradual on a timescale of months to years.  Between 1999 and
2001, the pulse morphology showed little or no change.

\begin{figure}[!htb]

\centerline{
\psfig{file=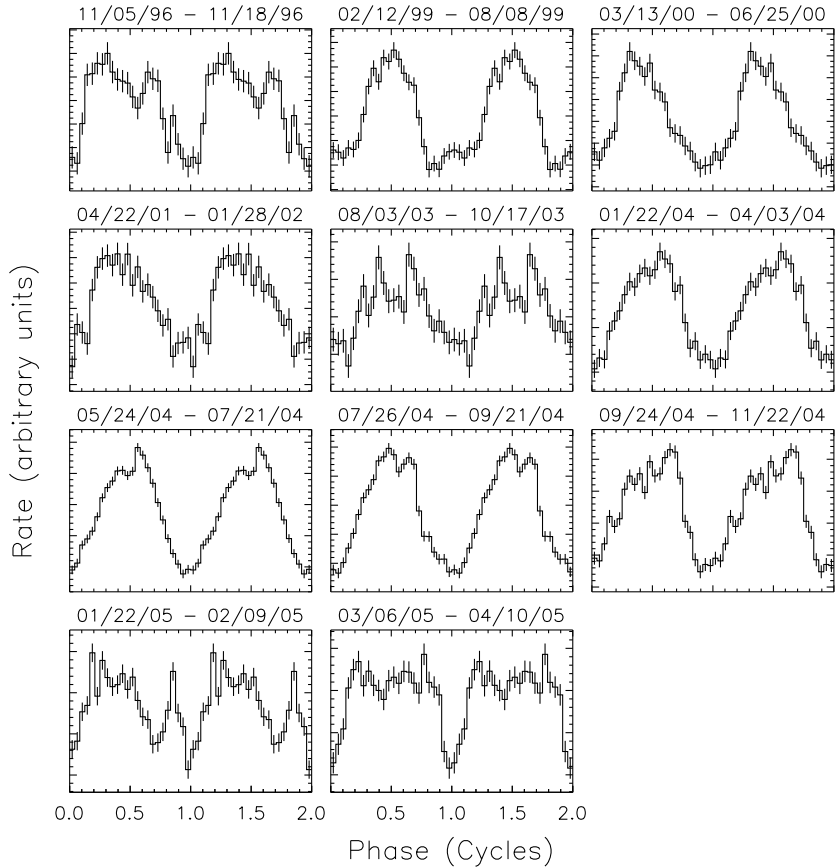,height=5.0in}}

\caption{Pulse morphology evolution of \sgra\ as seen with the \rxte\ PCA
between 1996 and 2005.  All pulse profiles shown are 2$-$10 keV and are
repeated once for clarity (0$-$2 cycles).  Note the change in pulse shape
across the giant flare from 2004 to 2005.  See text for details.
\label{fig:prof_time}}

\end{figure}

Folding our PCA data on the pulse ephemerides given in the last section, we
have extended the 2$-$10 keV pulse morphology history of \sgra\ through 2005
April (Figure~\ref{fig:prof_time}).  Very little additional change in pulse
shape was observed between 2001 and the months leading up to the giant flare. 
However, we note that there was one interval in 2003 where the profile was
temporarily more complex.  In the months preceding the flare, the source
brightened (see \S5) and the 2$-$10 keV pulse shape became somewhat more
jagged, yet retained the same overall pulse envelope.  The most profound change
occurred following the giant flare of 2004 December 27 when the pulse shape
exhibited two clear peaks in 2005 January/February, markedly different that the
pre-flare pulse shape over the same energy range.  However, this change appears
to have been short-lived as the pulse profile continued to evolve to a broad,
flat-topped peak in 2005 March/April.



Qualitatively similar pulse shape evolution was observed during the tail of the
giant flare from \sgra, albeit at much higher photon energies and luminosities
(e.g.\ Palmer et al.\ 2005).  Specifically, the complexity of the pulse profile
defined as the power contained in the higher harmonics relative to the
fundamental frequency {\it increased} during the tail of the flare.  Although
the direction of the pulse shape change in the quiescent emission was the same
(i.e.\ the persistent pulse shape became more complex following the flare), the
pulse shape of the persistent emission, even now, is much simpler than the
pulse shape at any time during the tail of the flare.

Flare-induced pulse shape changes have also been seen in \sgrc\ following the
1998 August 27 giant flare (Woods et al.\ 2001; {G\"o\u{g}\"u\c{s}} et al.\
2002).  In the case of \sgrc, the quiescent pulse profile suddenly changed from
a complex multi-peaked morphology before the giant flare to a nearly sinusoidal
single peak after the event.  Similarly, the pulse profile during the 5-minute
long flare tail evolved from a complex pulse pattern to a simpler, nearly
sinusoidal pulse shape toward the end.  Although both flares resulted in
sustained changes in the quiescent pulse shape, it is important to note that
the direction of the change was different for each flare.  The \sgrc\ pulse
profile simplified whereas the \sgra\ pulse profile became more complex.

\subsection{Energy Dependence}

{G\"o\u{g}\"u\c{s}} et al.\ (2002) noted that there was no significant energy
dependence of the pulse profile of \sgra\ over the energy range 2$-$30 keV
during PCA observations between 1996 and 2000.  The pulse profile during 2001
showed signs of greater complexity at high energies (20$-$30 keV), although the
signal-to-noise ratio for that data set was poor.  Similarly, the 2002 and 2003
data sets did not provide enough counts at energies above $\sim$7 keV to
construct meaningful pulse profiles.  Here, we investigate the energy
dependence of the \sgra\ pulse profile at epochs leading up to and following
the giant flare when the source was brightest.



\begin{figure}[!htb]

\centerline{
\psfig{file=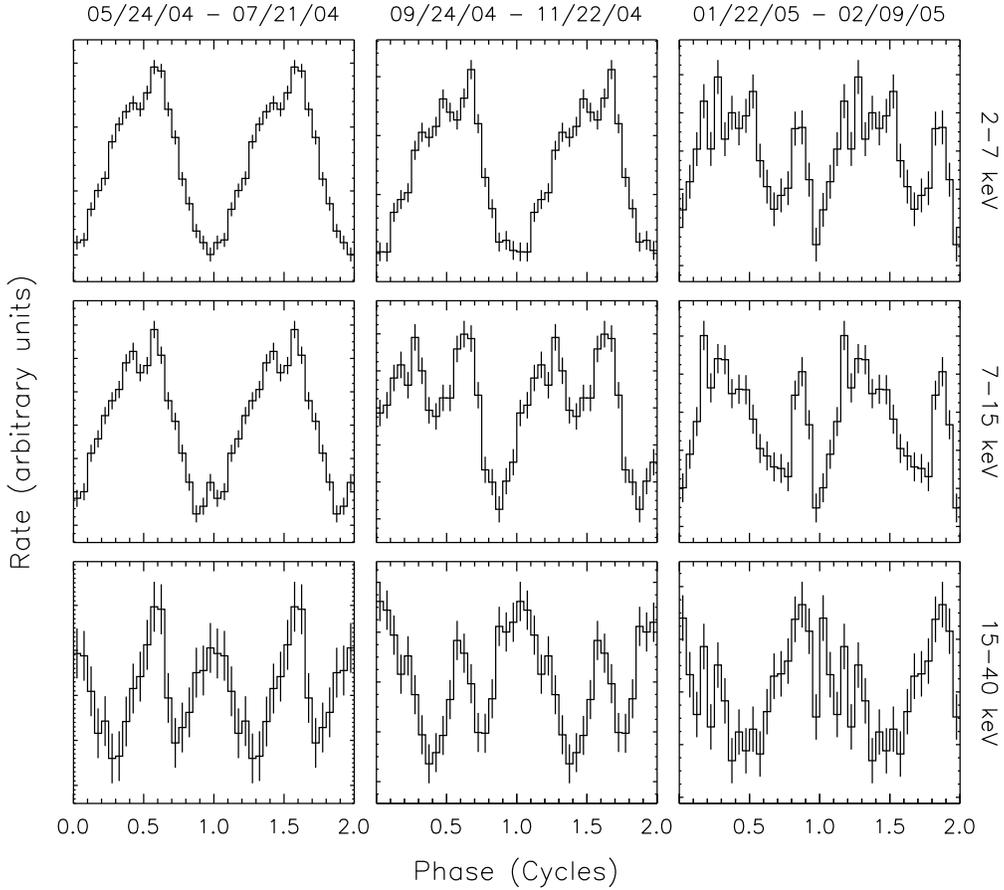,height=5.0in}}

\caption{Pulse profile evolution of \sgra\ in time and energy as seen with the
\rxte\ PCA in the months prior to and following the giant flare.  Time
increases from left to right and energy increases from top to bottom.  All
pulse profiles shown repeated once for clarity (0$-$2 cycles).  See text for
details. \label{fig:prof_energy}}

\end{figure}

Using the PCA data, we constructed three sets of pulse profiles over three
separate energy ranges between 2 and 40 keV (Figure~\ref{fig:prof_energy}). 
Approximately six months before the flare, the pulse profile below 15 keV was
fairly simple whereas the high energy profile (15$-$40 keV) showed two clear
peaks per rotation cycle.  The higher amplitude peak was correlated with the
much broader low-energy pulse maximum and the secondary peak was $\sim$0.5
cycle later in phase -- approximately aligned with pulse minimum at low
energies.  At two months prior to the flare, the pulse profile at intermediate
energies (7$-$15 keV) became two-peaked and the relative amplitudes of the two
peaks at high energies switched.  One month following the giant flare, the
pulse profile was very different showing multiple peaks at all energies.  The
dominant peak at high energies post-flare was seen as a narrow peak at
intermediate and low energies.  Although the most profound pulse shape changes
took place across the flare, it is clear that the pulse profile of \sgra\ was
evolving in both time and energy during the year leading up to the flare.

\subsection{Pulsed Fraction}

The pulsed fraction of \sgra\ is important in that it enables us to estimate
the total flux of the SGR when we do not know the precise level of the
background in the detector.  This is relevant for all \rxte\ PCA observations
which constitute the vast majority of our data set.  We estimate the total
(phase-averaged) flux of the SGR by taking the root mean square (r.m.s.) pulsed
flux and dividing by the r.m.s.\ pulsed fraction.  Here, we adopt the r.m.s.\
definition of the pulsed fraction given in Woods et al.\ (2004a).  The pulsed
flux is given by

\begin{equation}
F_{RMS} = \sqrt{\sum_{k=1}^{4} \frac{\alpha_{k}^{2} + \beta_{k}^{2} - 
   (\sigma_{\alpha_{k}}^{2} + \sigma_{\beta_{k}}^{2})}{2}},
\end{equation}

\noindent where

\begin{eqnarray*}
\alpha_{k} = \frac{2}{N} \sum_{i=1}^{N} r_i \cos{2\pi\phi_i{k}}, &
\displaystyle \beta_{k} = \frac{2}{N} \sum_{i=1}^{N} r_i \sin{2\pi\phi_i{k}}, \\
\sigma_{\alpha_{k}}^{2} = \frac{4}{N^2} \sum_{i=1}^{N} \sigma^{2}_{r_i} 
   \cos^2{2\pi\phi_i{k}},  &
\displaystyle \sigma_{\beta_{k}}^{2} = \frac{4}{N^2} \sum_{i=1}^{N} 
   \sigma^{2}_{r_i} \sin^2{2\pi\phi_i{k}}. \\
\end{eqnarray*}

\noindent Here, $F_{RMS}$ is the pulsed flux\footnote{Note that in Woods et
al.\ (2004) we used $F_{RMS}$ to denote pulsed fraction, not pulsed flux.},
$F_{RMS}/\bar{F}$ is the pulsed fraction where $\bar{F}$ is the phase-averaged
flux, $k$ refers to the harmonic number, $i$ refers to the phase bin, $N$ is
the total number of phase bins, $\phi_i$ is the phase, $r_i$ is the count rate
in the $i^{\rm th}$ phase bin, and $\sigma_{x_i}$ is the uncertainty in the
count rate of the $i^{\rm th}$ phase bin.  Note that the pulsed fractions
reported here may sometimes differ from measurements reported in the literature
by other authors using the same data sets.  These differences are due mostly to
differences in the definition of pulsed fraction.

When measuring the rms pulsed fraction, we used only data taken from X-ray
imaging telescopes where the background could be accurately measured.  For
consistency, we chose to measure the pulsed fraction over the energy range
2$-$10 keV.  For all observations, we extracted a source and background event
list for the given energy range, folded the source events on the measured pulse
period, subtracted the background count rate, and measured the rms pulsed
fraction using the sum of the first 4 harmonics.  The measured pulsed fractions
are plotted in Figure~\ref{fig:flux_hist}.  Prior to the giant flare, all
pulsed fractions of \sgra\ are consistent with being constant at $\sim$7\%. 
Following the giant flare there is a significant drop to 2.5$\pm$0.8\% during
the 2005 \cxo\ observation (Rea et al.\ 2005).  Subsequent \xmm\ observations
(Mereghetti et al.\ 2005) show a similarly low pulsed fraction, although more
recent observations show the pulsed fraction recovering to its pre-flare level
(Rea et al.\ 2005b).  We discuss the implications the changing pulsed fraction
has on our PCA pulsed flux normalization factor in \S5.4.

\section{X-ray Spectroscopy}

Up until recently, X-ray spectroscopic studies of the persistent,
phase-averaged emission from \sgra\ showed that the energy spectrum could be
modeled with a simple absorbed power-law (Sonobe et al.\ 1994; Mereghetti et
al.\ 2000).  Broad-band spectroscopy of the persistent emission has shown that
this non-thermal component extends up to at least 100 keV without signs of
rolling over (Mereghetti et al.\ 2005b; Molkov et al.\ 2005).  In the era of
high-throughput soft X-ray telescopes such as \cxo\ and \xmm, we are now able
to more precisely model the X-ray spectrum and identify deviations from the
simple power-law parameterization.

In general, magnetar candidates (i.e.\ SGRs and AXPs) have energy spectra that
are well modeled by the sum of a blackbody and a power-law.  From \cxo\
observations in 2004, we identified the long sought after blackbody component
in \sgra\ (Woods et al.\ 2004b).  Using \xmm\ observations, Mereghetti et al.\
(2005a) also identified a thermal component in the X-ray spectrum, however,
they measured a much higher temperature.  Here, we present our analysis of the
six \cxo\ observations of \sgra, place these results in context with the full
spectral history of this SGR, and address the apparent discrepancy between the
\cxo\ and \xmm\ blackbody temperatures.

\subsection{\cxo\ Spectral Analysis}

All six \cxo\ observations of \sgra\ were performed in CC mode having one
spatial dimension.  We extracted source spectra from within a 10 pixel
($5^{\prime \prime}$) region centered on the peak of the one-dimentional
image.  The background spectra were extracted from two 40 pixel ($20^{\prime
\prime}$) regions on either side of the SGR whose centers were offset from the
source centroid by 40 pixels.  As described in the section on \cxo\ timing
analysis (\S3.1), bursts were first removed from the event lists before
compiling the energy spectra.  The source spectra were rebinned to ensure at
least 25 counts were contained within each energy bin.  The effective area
files and response matrices were constructed using the CIAO 3.2.1 procedures
{\tt mkrmf} and {\tt mkarf}, respectively.  The calibration database used to
create these files was version 3.0.1.

Using XSPEC v11.3, we fit each spectrum individually to a power-law (PL) model
and the sum of a blackbody plus a power law (BB$+$PL).  A narrow feature at 1.7
keV was seen in the residuals of all fits.  This feature is almost certainly
instrumental in origin due to inaccuracies of the instrumental response. 
Artificial narrow features between 1.5 and 2.0 keV are commonly observed in
\cxo\ CC-mode energy spectra of bright point sources.  For this reason, we have
limited our fit range to energy channels where the response is best calibrated
between 0.5$-$1.6 and 2.0$-$10.0 keV.

For all six observations, the $\chi^2$ improved when the blackbody component
was included in the fit.  The improvement in $\chi^2$ varied from 7 to 19 with
an average change in $\chi^2$ of 14.  The significance of the thermal component
in the observed spectrum was, on average, marginal for any given data set.  To
more sensitively probe our model comparison, we fit all six spectra
simultaneously, forcing only the column density to be linked for all spectra. 
Comparing the PL and BB$+$PL model fits, we found that the total $\chi^2$
dropped by 93 with the addition of the 12 free blackbody parameters in the
simultaneous fit.  The F-test between these two models yielded a probability of
4 $\times$ 10$^{-14}$, indicating that the BB$+$PL model was strongly favored
over the simple PL model.  All fit parameters for both the PL and BB$+$PL
models are listed in Table~\ref{tab:cxo_spec}.


\begin{table}[!htb]
\begin{minipage}{1.0\textwidth}
\begin{small}
\begin{center}
\caption{Measured spectral parameters of \sgra\ from \cxo\ 
observations. \label{tab:cxo_spec}} 
\vspace{10pt}
\begin{tabular}{ccccccc} \hline \hline

 Observation & Model$^{a}$ & N$_{H}$ & $kT$ & $\Gamma$ & Flux & Unabs.\ Flux \\
  &  & (10$^{22}$ cm$^{-2}$) & (keV) & & \multicolumn{2}{c}{(10$^{-11}$ ergs cm$^{-2}$ s$^{-1}$)$^{b}$}  \\\hline
 
 CXO1 &   PL       &  7.1(3)  &  ...    & 1.89(8)  & 1.21 & 1.95  \\
      & BB$+$PL    &  8.0(8)  & 0.48(9) & 1.57(23) & 1.27 & 2.20  \\
      &   PL(s)    &  7.88(7) &  ...    & 2.09(4)  & 1.17 & 2.05  \\
      & BB$+$PL(s) &  8.78(3) & 0.41(4) & 1.69(14) & 1.26 & 2.39  \\
      & BB$+$PL(us)&  7.19(12)& 0.57(5) & 1.40(20) & 1.28 & 2.05  \\

 CXO2 &   PL       &  8.1(2)  &  ...    & 1.94(5)  & 1.88 & 3.19  \\
      & BB$+$PL    &  8.6(5)  & 0.49(8) & 1.70(14) & 1.94 & 3.43  \\
      &   PL(s)    &  7.88(7) &  ...    & 1.89(3)  & 1.89 & 3.16  \\
      & BB$+$PL(s) &  8.78(3) & 0.45(4) & 1.76(10) & 1.93 & 3.50  \\
      & BB$+$PL(us)&  7.19(12)& 0.75(3) & 1.14(17) & 1.98 & 3.06  \\

 CXO3 &   PL       &  7.7(3)  &  ...    & 1.77(7)  & 2.09 & 3.38  \\
      & BB$+$PL    &  8.5(8)  & 0.49(9) & 1.47(20) & 2.18 & 3.76  \\
      &   PL(s)    &  7.88(7) &  ...    & 1.81(4)  & 2.08 & 3.41  \\
      & BB$+$PL(s) &  8.78(3) & 0.47(5) & 1.50(14) & 2.18 & 3.86  \\
      & BB$+$PL(us)&  7.19(12)& 0.71(4) & 0.98(24) & 2.23 & 3.39  \\

 CXO4 &   PL       &  7.7(2)  &  ...    & 1.64(5)  & 2.39 & 3.75  \\
      & BB$+$PL    &  7.8(5)  & 0.59(14)& 1.39(19) & 2.45 & 3.87  \\
      &   PL(s)    &  7.88(7) &  ...    & 1.69(3)  & 2.37 & 3.79  \\
      & BB$+$PL(s) &  8.78(3) & 0.43(4) & 1.56(9)  & 2.42 & 4.20  \\
      & BB$+$PL(us)&  7.19(12)& 0.75(5) & 1.14(16) & 2.48 & 3.70  \\

 CXO5 &   PL       &  8.1(2)  &  ...    & 1.85(5)  & 2.47 & 4.13  \\
      & BB$+$PL    &  9.0(6)  & 0.44(6) & 1.69(12) & 2.53 & 4.61  \\
      &   PL(s)    &  7.88(7) &  ...    & 1.79(3)  & 2.49 & 4.08  \\
      & BB$+$PL(s) &  8.78(3) & 0.46(5) & 1.67(10) & 2.53 & 4.50  \\
      & BB$+$PL(us)&  7.19(12)& 0.78(4) & 1.11(17) & 2.59 & 3.94  \\

 CXO6 &   PL       &  8.0(2)  &  ...    & 2.06(6)  & 2.05 & 3.57  \\
      & BB$+$PL    & 10.2(8)  & 0.33(3) & 2.09(12) & 2.07 & 4.74  \\
      &   PL(s)    &  7.88(7) &  ...    & 2.04(4)  & 2.06 & 3.55  \\
      & BB$+$PL(s) &  8.78(3) & 0.41(4) & 1.91(10) & 2.10 & 3.98  \\
      & BB$+$PL(us)&  7.19(12)& 0.70(4) & 1.37(18) & 2.16 & 3.43  \\

\hline\hline
\end{tabular}
\end{center}
\noindent$^{a}$ PL = Power law; BB$+$PL = Blackbody plus power law; (s) 
indicates a simultaneous fit with the column density linked between all \cxo\
observations; (us) indicates a universal simultaneous fit with the column 
density linked between all observations (\cxo, \xmm, \sax\ and \asca). \\
\noindent$^{b}$ Integrated over the energy range 2$-$10 keV. 
\end{small}
\end{minipage}\hfill
\end{table}

The average blackbody temperature of \sgra\ measured using the \cxo\ data is
0.44 keV, very near the measured temperature of \sgrc\ as well as most other
magnetar candidates (e.g.\ Woods \& Thompson 2006).  However, we find that the
temperature we measure is systematically smaller than the temperature measured
using \xmm\ data (Mereghetti et al.\ 2005a), even when the \cxo\ and \xmm\
observations are nearly simultaneous.  For example, \xmm\ observed \sgra\ on
2004 October 6 (ObsD in Mereghetti et al.), just 3 days before CXO5 with \cxo. 
For this \xmm\ observation, Mereghetti et al.\ (2005) measured a temperature of
0.77$\pm$0.15 keV, a photon index of $-$1.2$\pm$0.2, and a column density of
6.5$\pm$0.6 $\times$ 10$^{22}$ cm$^{-2}$, all significantly different than the
parameters derived from the CXO5 \cxo\ data (see Table~\ref{tab:cxo_spec}).  In
an effort to resolve this discrepancy, we analyzed this observation and all
other publicly available \xmm\ observations of \sgra.

\subsection{Comparison to \xmm\ Results}

There have been six observations of \sgra\ carried out with \xmm\ between 2003
April and 2005 October.  The times of these observations and their approximate
exposure times are listed in Table~\ref{tab:xmmobs}.  An analysis of the four
observations through 2004 October has been presented in Mereghetti et al.\
(2005a).  The post-flare \xmm\ observations of \sgra\ were presented by Tiengo
et al.\ (2005) and Rea et al.\ (2005b).  Here, we present our analysis of PN
energy spectra of the persistent X-ray emission recorded during all six \xmm\
observations.


\begin{table}[!h]
\begin{minipage}{1.0\textwidth}
\begin{center}
\caption{\xmm\ observation log for \sgra\ between 2003 April and 2005 October. 
\label{tab:xmmobs}} 
\vspace{10pt}
\begin{tabular}{cccc} \hline \hline

 Name &     \xmm\      &     Date     & PN Exposure  \\
      & Observation ID &              &   (ks)    \\\hline
 
 XMM1 &  0148210101    & 2003 Apr 03  &   55.5    \\
 XMM2 &  0148210401    & 2003 Oct 07  &   22.4    \\
 XMM3 &  0205350101    & 2004 Sep 06  &   51.9    \\
 XMM4 &  0164561101    & 2004 Oct 06  &   18.9    \\
 XMM5 &  0164561301    & 2005 Mar 07  &   24.9    \\
 XMM6 &  0164561401    & 2005 Oct 04  &   33.0    \\

\hline\hline
\end{tabular}
\end{center}
\end{minipage}\hfill
\end{table}

During the first two \xmm\ observations of \sgra\ in 2003, the PN camera was
operated in Full Frame (FF) mode.  The four subsequent observations have been
operated in Small Window (SW) mode to better study SGR burst emissions.  The SW
mode has finer time resolution and can tolerate a greater dynamic flux range
than FF mode.  In all observations, the medium thickness filter was used. 
Starting from the Observation Data Files, we processed the PN data using the
script {\tt epchain} provided in the {\it XMM} Science Analysis Software
(XMMSAS) v6.5.0.  Next, we constructed a light curve of the central PN CCD,
excluding the bright central source, and identified times of high background. 
We chose a threshold of 2 times the nominal 0.5$-$10.0 keV background to define
regions of high background.  Accordingly, we filtered out 0$-$40\% of the total
exposure from each data set before subsequent analysis.  Finally, we
constructed light curves of \sgra\ at 1 s time resolution to identify bursts. 
Using custom software, we filtered out several tens of SGR bursts from the
event lists.

Using our filtered event lists, we extracted source spectra from $37.5^{\prime
\prime}$ (750 pixel) radii circular regions centered on the SGR and background
spectra from $\sim67^{\prime \prime}$ radii circular regions from the same
CCD.  We followed standard XMMSAS recipes in grade selection (pattern $<$4) and
generation of effective area files and response matrices.

Using XSPEC v11.3, we fit the individual \xmm\ spectra over the energy range
0.5$-$10.0 keV to both the PL and BB$+$PL models.  Similar to the \cxo\
spectral results, we measured small changes in $\chi^2$ for 4 individual
spectral fits ($\Delta \chi^2 = 9-18$).  The two exceptions were observations
XMM3 and XMM5 which yielded a reduction in $\chi^2$ of 27 and 37, respectively,
between the PL and BB$+$PL models.  The improvement in $\chi^2$ for these two
data sets was significant.  The combined simultaneous fit to all \xmm\ energy
spectra indicated that the inclusion of the blackbody component was again very
significant (F-test probability $\approx 10^{-14}$).  The fit parameters for
all spectral fits are given in Table~\ref{tab:xmm_spec}.  We note that the fit
parameters we measure are mostly consistent with the results of Mereghetti et
al.\ (2005a), Tiengo et al.\ (2005), and Rea et al.\ (2005b).  On average, we
measure slightly higher column densities and steeper photon indices than
Mereghetti et al..  These subtle differences could be caused by choices of
energy fit range and/or binning -- for example.


\begin{table}[!htb]
\begin{minipage}{1.0\textwidth}
\begin{small}
\begin{center}
\caption{Measured spectral parameters of \sgra\ from \xmm\
observations. \label{tab:xmm_spec}} 
\vspace{10pt}
\begin{tabular}{ccccccc} \hline \hline

 Observation & Model$^{a}$ & N$_{H}$ & $kT$ & $\Gamma$ & Flux & Unabs.\ Flux \\
  &  & (10$^{22}$ cm$^{-2}$) & (keV) & & \multicolumn{2}{c}{(10$^{-11}$ ergs cm$^{-2}$ s$^{-1}$)$^{b}$}  \\\hline
 
 XMM1 &   PL       &  6.6(3)  &  ...    & 1.63(6)  & 1.08 & 1.61  \\
      & BB$+$PL    &  7.2(7)  & 0.54(12)& 1.41(15) & 1.09 & 1.72  \\
      &   PL(s)    &  7.12(6) &  ...    & 1.73(4)  & 1.07 & 1.67  \\
      & BB$+$PL(s) &  6.63(2) & 0.65(7) & 1.29(15) & 1.10 & 1.64  \\
      & BB$+$PL(us)&  7.19(12)& 0.54(6) & 1.41(12) & 1.09 & 1.71  \\

 XMM2 &   PL       &  6.7(2)  &  ...    & 1.64(5)  & 1.20 & 1.80  \\
      & BB$+$PL    &  6.8(5)  & 0.66(12)& 1.32(18) & 1.21 & 1.84  \\
      &   PL(s)    &  7.12(6) &  ...    & 1.72(3)  & 1.19 & 1.85  \\
      & BB$+$PL(s) &  6.63(2) & 0.72(5) & 1.16(14) & 1.22 & 1.81  \\
      & BB$+$PL(us)&  7.19(12)& 0.61(5) & 1.32(11) & 1.21 & 1.90  \\

 XMM3 &   PL       &  7.2(1)  &  ...    & 1.56(2)  & 2.48 & 3.75  \\
      & BB$+$PL    &  6.7(3)  & 0.85(7) & 1.14(14) & 2.50 & 3.63  \\
      &   PL(s)    &  7.12(6) &  ...    & 1.55(2)  & 2.48 & 3.74  \\
      & BB$+$PL(s) &  6.63(2) & 0.86(5) & 1.12(10) & 2.50 & 3.62  \\
      & BB$+$PL(us)&  7.19(12)& 0.71(5) & 1.34(7)  & 2.50 & 3.78  \\

 XMM4 &   PL       &  7.3(2)  &  ...    & 1.69(4)  & 2.44 & 3.80  \\
      & BB$+$PL    &  6.7(5)  & 0.84(10)& 1.21(24) & 2.46 & 3.64  \\
      &   PL(s)    &  7.12(6) &  ...    & 1.65(2)  & 2.44 & 3.75  \\
      & BB$+$PL(s) &  6.63(2) & 0.85(5) & 1.18(13) & 2.46 & 3.62  \\
      & BB$+$PL(us)&  7.19(12)& 0.72(5) & 1.42(10) & 2.45 & 3.78  \\

 XMM5 &   PL       &  7.1(2)  &  ...    & 1.72(4)  & 1.95 & 3.03  \\
      & BB$+$PL    &  6.0(4)  & 0.91(6) & 0.65(36) & 1.99 & 2.80  \\
      &   PL(s)    &  7.12(6) &  ...    & 1.72(3)  & 1.95 & 3.04  \\
      & BB$+$PL(s) &  6.63(2) & 0.79(4) & 1.05(16) & 1.98 & 2.95  \\
      & BB$+$PL(us)&  7.19(12)& 0.70(4) & 1.27(12) & 1.98 & 3.09  \\

 XMM6 &   PL       &  7.2(2)  &  ...    & 1.83(4)  & 1.30 & 2.08  \\
      & BB$+$PL    &  6.6(4)  & 0.77(7) & 1.19(22) & 1.32 & 2.00  \\
      &   PL(s)    &  7.12(6) &  ...    & 1.81(3)  & 1.54 & 2.07  \\
      & BB$+$PL(s) &  6.63(2) & 0.76(4) & 1.28(13) & 1.32 & 2.00  \\
      & BB$+$PL(us)&  7.19(12)& 0.65(4) & 1.49(10) & 1.32 & 2.10  \\

\hline\hline
\end{tabular}
\end{center}
\noindent$^{a}$ PL = Power law; BB$+$PL = Blackbody plus power law; (s) 
indicates simultaneous fit with the column density linked between all
\xmm\ observations; (us) indicates a universal simultaneous fit with the column 
density linked between all observations (\cxo, \xmm, \sax\ and \asca).   \\
\noindent$^{b}$ Integrated over the energy range 2$-$10 keV. 
\end{small}
\end{minipage}\hfill
\end{table}

When plotted on the same scale, {\it all} \xmm\ spectral measurements
(including blackbody temperature) resulting from individual spectral fits are
systematically offset from nearby \cxo\ measurements indicating a discrepancy
between the two instruments.  The consistent offset in individual spectral
parameters suggests that the differences are instrumental and not due to
intrinsic variability of the SGR.

In spite of the differences between the \cxo\ and \xmm\ spectral parameters,
our joint analysis of the two data sets allowed us to conclude that ($i$) the
simple power-law model does not accurately represent the X-ray energy spectrum
of \sgra\ and ($ii$) the addition of a thermal component yields acceptable
spectral fits.  To further investigate the residual differences between the
\cxo\ and \xmm\ results, we attempted inter-instrument simultaneous fitting of
all available \sgra\ data sets.

\subsection{Universal Simultaneous Fit}

Prior to the 12 \cxo\ and \xmm\ observations of \sgra\ presented here, there
were four \sax\ observations between 1998 and 2001 (Mereghetti et al.\ 2000)
and two \asca\ observations in 1993 (Sonobe et al.\ 1994) and 1995 suitable for
spectral fitting.  The \asca\ GIS data were processed following standard
analysis procedures outlined in the \asca\ data analysis
guide\footnote{http://heasarc.gsfc.nasa.gov/docs/asca/abc/abc.html}. 
Similarly, the \sax\ LECS and MECS data were processed using {\tt Xselect} as
directed in the \sax\ 
guide\footnote{http://heasarc.gsfc.nasa.gov/docs/sax/abc/saxabc.html}.  As with
the previous simultaneous fits, we forced the column density to be the same for
all observations.  Due to poorer signal-to-noise quality of the \asca\ spectra,
we fixed the blackbody temperature for these data sets equal to the mean of the
four measured \sax\ temperatures.  All other spectral parameters were free to
vary in the fit.  The measured values are listed in Tables~\ref{tab:cxo_spec},
\ref{tab:xmm_spec} and \ref{tab:oth_spec} in the rows labeled ``us.''


\begin{table}[!htb]
\begin{minipage}{1.0\textwidth}
\begin{small}
\begin{center}
\caption{Measured spectral parameters of \sgra\ from \asca\ and \sax\
observations. \label{tab:oth_spec}} 
\vspace{10pt}
\begin{tabular}{ccccccc} \hline \hline

 Observation & Model$^{a}$ & N$_{H}$ & $kT$ & $\Gamma$ & Flux & Unabs.\ Flux \\
  &  & (10$^{22}$ cm$^{-2}$) & (keV) & & \multicolumn{2}{c}{(10$^{-11}$ ergs cm$^{-2}$ s$^{-1}$)$^{b}$}  \\\hline
 
ASCA1 & BB$+$PL(us) &  7.19(12) & 0.476    & 1.44(13) & 0.91 & 1.70  \\

ASCA2 & BB$+$PL(us) &  7.19(12) & 0.476    & 1.67(15) & 0.70 & 1.36  \\

 SAX1 & BB$+$PL(us) &  7.19(12) & 0.49(8)  & 1.75(22) & 1.04 & 1.79  \\

 SAX2 & BB$+$PL(us) &  7.19(12) & 0.44(10) & 1.95(13) & 1.06 & 1.83  \\

 SAX3 & BB$+$PL(us) &  7.19(12) & 0.47(7)  & 1.77(20) & 0.99 & 1.73  \\

 SAX4 & BB$+$PL(us) &  7.19(12) & 0.50(6)  & 1.66(16) & 1.15 & 1.96  \\

\hline\hline
\end{tabular}
\end{center}
\noindent$^{a}$ BB$+$PL = Blackbody plus power law; (us) indicates a universal 
simultaneous fit with the column density linked between all observations (\cxo, 
\xmm, \sax\ and \asca).   \\
\noindent$^{b}$ Integrated over the energy range 2$-$10 keV. 
\end{small}
\end{minipage}\hfill
\end{table}

Simultaneously fitting all available \sgra\ data with a single column density
fitted for all observations significantly reduced the discrepancy between our
\cxo\ results and the \xmm\ results.  For the universal simultaneous fit, we
obtain a statistically acceptable $\chi^2$ of 8379 for 8421 degrees of
freedom.  We now find very good agreement between the measured blackbody
temperatures, power-law photon indices, and X-ray fluxes.  For example,
consider the near simultaneous \cxo\ and \xmm\ observations in 2004 October
(CXO5 and XMM4).  For the independent spectral fits, the measured blackbody
temperatures differed by 3.5$\sigma$, the photon index by 1.8$\sigma$, and
unabsorbed flux by 26\%.  When we linked the column density in the universal
simultaneous fit, these differences were reduced to less than 1.6$\sigma$ for
all parameters (see Table~\ref{tab:cxo_spec} and \ref{tab:xmm_spec} for
additional examples).

The improved agreement between the \cxo\ and \xmm\ results for the simultaneous
fit with a linked column density suggests that the instrumental ``discrepancy''
we noted originally is likely due to strong coupling of the spectral parameters
in combination with slight differences in the instrumental response functions
of the two instruments.  The cross-correlation between the blackbody parameters
and the column density is particularly strong and that is where we observed the
largest disparity.  By forcing the column density to be the same for all data
sets, we effectively reduced the covariance between these parameters.

\subsection{Spectral History}

\begin{figure}[!p]

\centerline{
\psfig{file=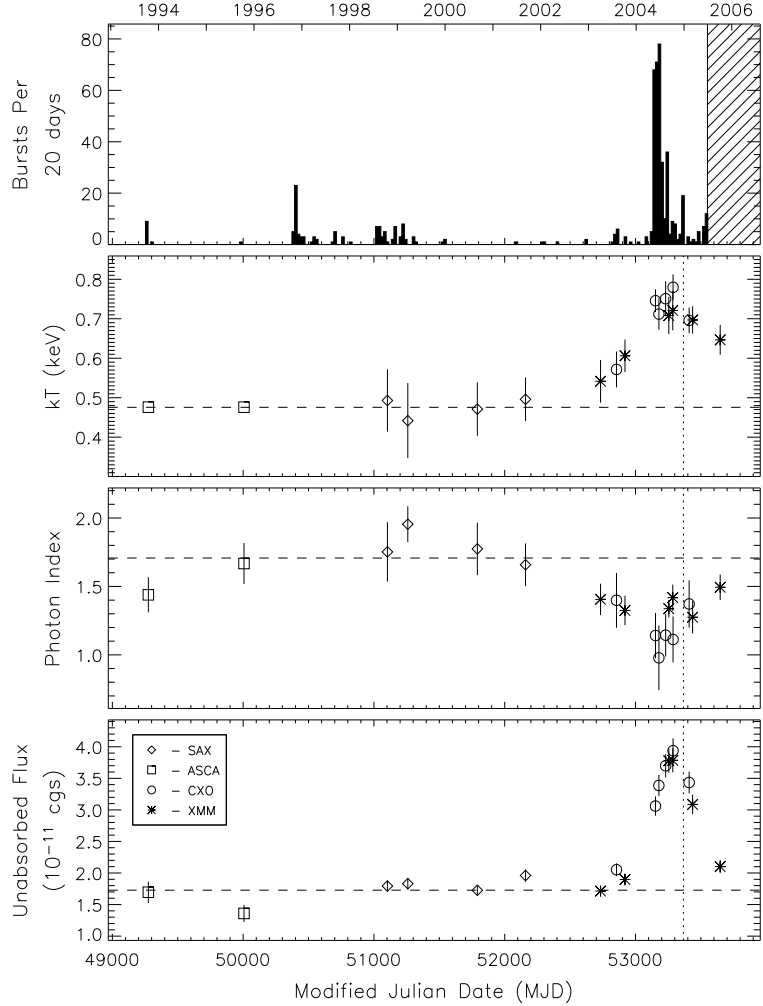,height=6.0in}}

\caption{Spectral history of \sgra\ from 1993 to 2005 using the blackbody plus
power-law model simultaneously fit to all data sets.  The measured column
density for this fit was 7.19$\pm$0.12 cm$^{-2}$.  From top to bottom, IPN
burst rate history, blackbody temperature, photon index, and unabsorbed flux. 
The time of the giant flare is indicated by a vertical dotted line.  The burst
rate data are complete through 2005 June.  Horizontal dashed lines indicate the
average value for the given parameter for measurements before 2002.  See text
for details. \label{fig:spec_hist}}

\end{figure}

Combining our \cxo, \xmm, \sax, and \asca\ spectral results on \sgra, we
constructed a comprehensive spectral history of the SGR from 1993 to 2005
(Figure~\ref{fig:spec_hist}).  Shown are the spectral parameters derived from
the universal simultaneous spectral fit described in the previous section. 
Note that the blackbody temperature was fixed for the \asca\ spectra to the
average of the four \sax\ temperatures.  As can be seen in this figure, the
unabsorbed flux showed very little variability between 1993 and 2002 before
increasing by more than a factor 2 during the 2004 burst active episode. 
Correlated with the peak in flux in 2004 was a maximum in blackbody temperature
and minimum in photon index.  The increased spectral hardness was evidenced in
both the thermal and non-thermal components of the spectrum.  Interestingly,
each began to show changes in early 2003 -- more than one year prior to the
giant flare (vertical dotted line).

As with the torque on the star, the peaks (valley) in these three spectral
parameters appear to precede the flare itself.  We fitted the blackbody
temperature and photon index measurements between MJD 52700 and 53700 to a
quadradic and measured centroids of 53280$\pm$40 and 53160$\pm$60,
respectively.  The X-ray flux was more peaked, so we limited our fit range to
MJD 53000 to 53500 and measured a centroid of 53296$\pm$8.  All three cenroid
values precede the giant flare (MJD 53366) by several months.  However, our
data coverage for spectral measurements is admittedly much more sparse than our
frequency derivative measurements and these maxima are relatively broad.



\begin{figure}[!p]

\centerline{
\psfig{file=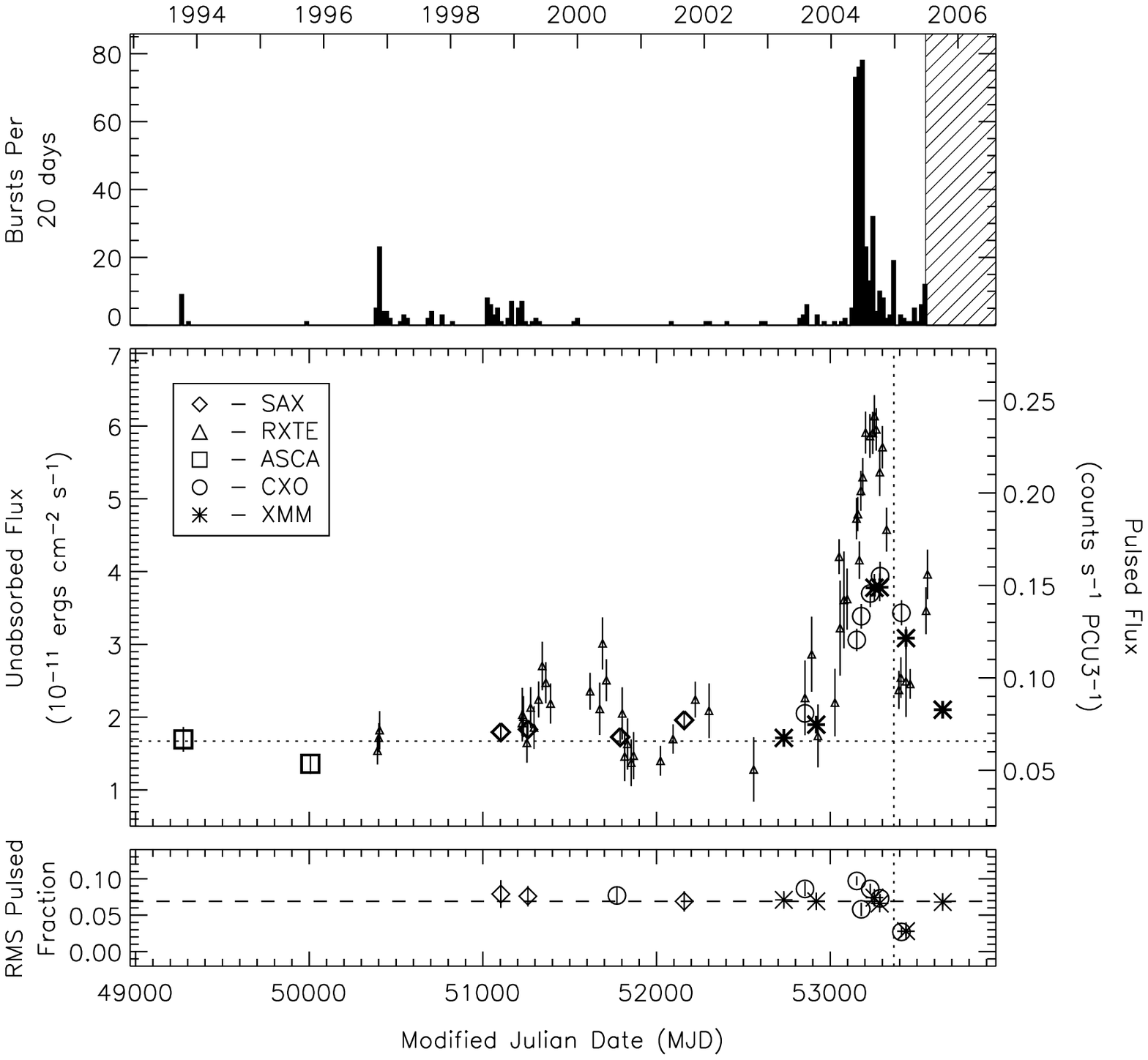,height=6.0in}}

\caption{Persistent and pulsed flux history of \sgra\ between 1993 and 2005.
{\it Top} -- Burst rate history (through 2004 October) as seen with instruments
within the Inter-Planetary Network.  The time of the giant flare is indicated
in subsequent panels by a vertical dotted line.  The burst rate data are
complete through 2005 June.  {\it Middle} -- Persistent and pulsed flux history
of the SGR (2$-$10 keV).  Unabsorbed fluxes (left axis) measured using imaging
X-ray telescopes such as \asca, \sax, \cxo, and \xmm.  The pulsed fluxes (right
axis) are measured using \rxte.  The pulsed flux measurements are
``normalized'' to the phase-averaged flux level assuming a constant pulsed
fraction of $\sim$7\%.  The horizontal dotted line indicates the pre-2000
average flux level using ASCA and SAX measurements.  Solid curved lines
indicate fits to a quadratic model for data between MJD 53000 and 53500.  See
text for details.  {\it Bottom} -- Pulsed fraction (2$-$10 keV) as measured
using the imaging X-ray telescopes. \label{fig:flux_hist}}

\end{figure}

To further investigate the flux variability of \sgra, we included pulsed flux
measurements of the SGR obtained using \rxte\ PCA data.  Following the method
described in Woods et al.\ (2001 and 2004) for \sgrc\ and \axpa, respectively,
we folded individual segments of 2$-$10 keV PCA data to created high
signal-to-noise pulse profiles.  We computed the r.m.s.\ pulsed amplitude of
each segment by summing the power of the first 4 harmonics according to
equation~1.  In Figure~\ref{fig:flux_hist}, we show the pulsed flux and
phase-averaged unabsorbed flux values (also plotted in
Figure~\ref{fig:spec_hist}).  The far more numerous PCA pulsed fluxes provide a
more comprehensive picture of the flux evolution of the SGR over the last
decade.  The pulsed flux axis (right) is referenced to the phase-averaged flux
axis by calculating a scale factor between the two from PCA pulsed flux
measurements in 1999 and a contemporaneous phase-averaged flux measurement from
\sax.  Assuming that the pulsed fraction of \sgra\ remains constant (and
perfect X-ray detector intercalibration), the PCA pulsed fluxes on this scale
would exactly match all other phase-averaged fluxes.  With the exception of the
months leading up to and following the giant flare, there is generally good
agreement between the two.  The post-flare disparity is clearly due to the
sudden drop in pulsed fraction (bottom panel).  The pre-flare mismatch could be
due to a change in the energy dependence of the pulsed fraction during the flux
rise.



\begin{figure}[!p]

\centerline{
\psfig{file=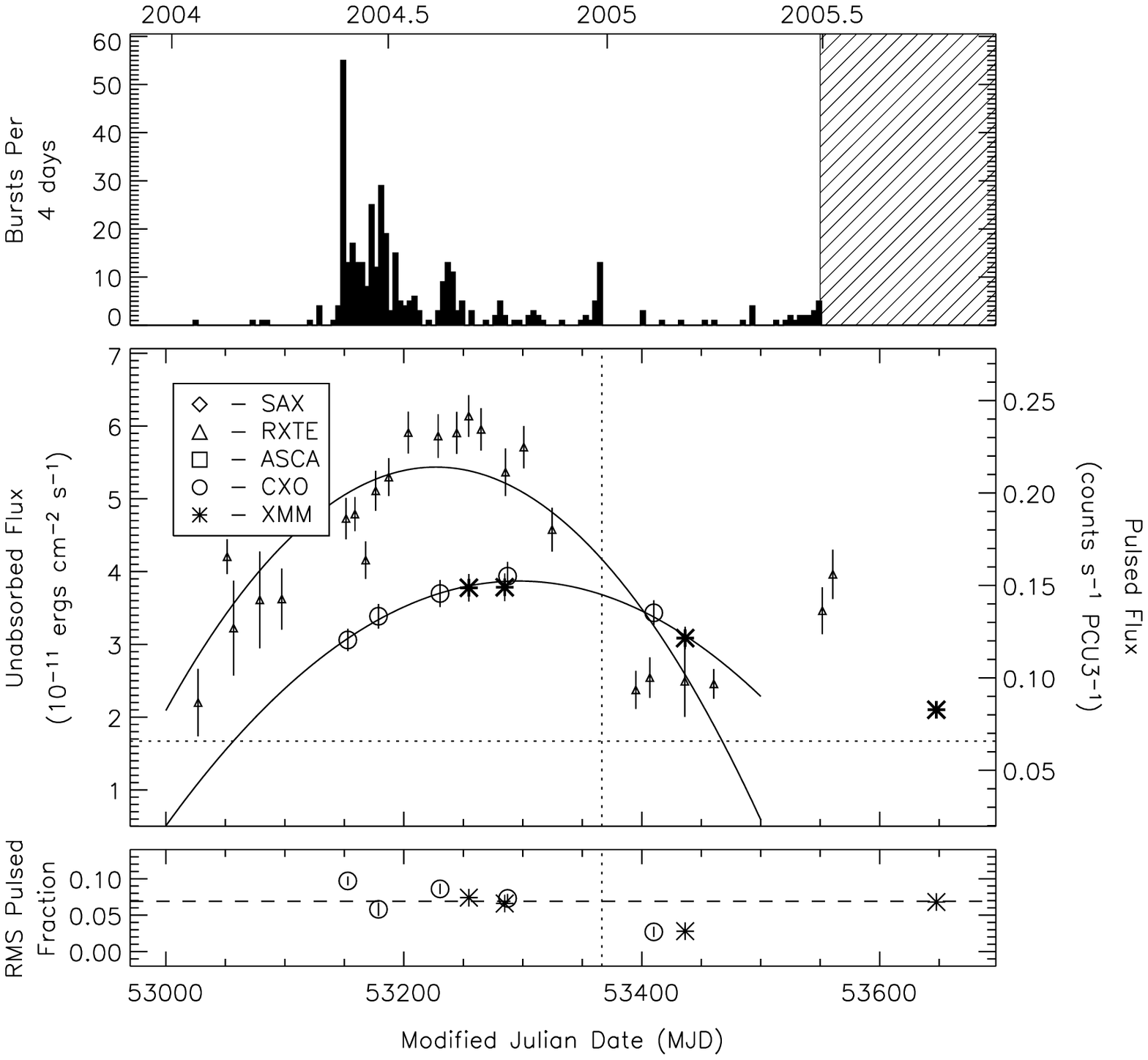,height=6.0in}}

\caption{Persistent and pulsed flux history of \sgra\ between 2004 and 2005. 
{\it Top} -- Burst rate history (through 2004 October) as seen with instruments
within the Inter-Planetary Network.  The time of the giant flare is indicated
in subsequent panels by a vertical dotted line.  The burst rate data are
complete through 2005 June.  {\it Middle} -- Persistent and pulsed flux history
of the SGR (2$-$10 keV).  Unabsorbed fluxes (left axis) measured using imaging
X-ray telescopes such as \cxo\ and \xmm.  The pulsed fluxes (right axis) are
measured using \rxte.  The pulsed flux measurements are ``normalized'' to the
phase-averaged flux level assuming a constant pulsed fraction of 7\%.  See text
for details.  {\it Bottom} -- Pulsed fraction (2$-$10 keV) as measured using
the imaging X-ray telescopes. \label{fig:flux_hist_flare}}

\end{figure}

Low-level changes in the pulsed flux of \sgra\ are evident between 1999 and
2003, although the largest magnitude changes in flux occurred during the time
leading up to and following the giant flare.  A close-up of the flux evolution
during this epoch (Figure~\ref{fig:flux_hist_flare}) shows that the flux rose
on a timescale of months in the build-up to the flare.  As with the torque,
spectral hardness and phase-averaged flux, the pulsed flux peaks well before
the flare itself on 2004 December 27.  Fitting the pulsed flux data between MJD
53000 and 53500 to a quadratic, we find the centroid at MJD 53227$\pm$8, nearly
5 months prior to the flare.



\section{Discussion}

Similar to outbursts in other magnetar candidates, the intense burst activity
of \sgra\ in 2004 was accompanied by changes in the persistent and pulsed
emission properties of the source.  Specifically, we observed a hardening of
the X-ray spectrum, large amplitude increases in the pulsed and phase-averaged
flux, strong variability in the spin-down rate, and significant changes in the
pulse morphology.  The connection between burst activity and the persistent
emission of magnetar candidates has allowed us to place constraints on the
magnetar model, and at times motivate refinements to the model.  Below, we
discuss how the changes we observed in the persistent X-ray emission of \sgra\
during the 2004 outburst fit within the magnetar model.

Mereghetti et al.\ (2005a) reported a correlation between spectral hardness and
spin-down rate (i.e.\ torque) of \sgra\ for X-ray observations preceding the
giant flare.  They report a monotonic increase in spectral hardness from 1993
to 2004.  Our analysis of the data is consistent with their conclusion that the
energy spectrum hardens over the last decade, however, we find significant
differences in the temporal evolution and consequently the hardness-torque
correlation.  With regard to the spectral evolution, our universal simultaneous
fit shows that the X-ray spectrum does not begin to harden until at least 1999.
In fact, the data do not show a clear hardening trend until 2003. From this
period through mid-2005, the photon index steadily flattens with time and the
blackbody temperature increases.  Our \rxte\ monitoring observations of \sgra\
show that the torque change, on the other hand, was relatively sudden in 2000
taking $\sim$1 year to transition to a new equilibrium state where the
spin-down rate remained roughly constant until 2004.  Only during this $\sim$4
year interval of steady, enhanced torque did the energy spectrum become
harder.  If, in fact, the two effects are correlated in \sgra, the torque
change in year 2000 preceded the gradual hardening of the spectrum as opposed
to a monotonic evolution of each parameter in lock-step with the other as
suggested by Mereghetti et al..

Early in 2004, the torque on \sgra\ began to increase again reaching a maximum
$\sim$2 months after the peak in burst rate, but still several months before
the flare epoch.  Similarly, the energy spectrum peaked in hardness after the
burst rate peak and before the flare epoch.  The spectral hardness peak appears
to be delayed relative to the torque maximum.  As the torque underwent a rapid
decline, the energy spectrum followed with a gradual softening.  These trends
continued through the flare epoch without deviating.  Approximately three
months after the flare, the torque reached a local minimum and has since
recovered to the pre-flare level of 2001$-$2004.  The spectral hardness, on the
other hand, has continued to drop and is steadily approaching the pre-2000
spectral shape.

In summary, the correlation between torque and spectral hardness is not
straightforward.  There is evidence in favor of the torque change in year 2000
preceding, perhaps triggering a gradual hardening of the energy spectrum. 
However, the pre-flare drop in torque and its subsequent post-flare recovery
resulted in a spectral softening, but no recovery (as of yet) in the spectral
hardness.  This may indicate some level of histeresis in the system.

A correlation between spectral hardness and torque is expected in the model of
Thompson et al.\ (2002) for a magnetar with a twisted magnetosphere.  It is
hypothesized that a twist is imparted on the magnetosphere from below as
residual magnetic field complexities within the interior of the star work their
way to the surface and deform the crust in energetically favorable rotational
motions.  The subsequent twisting of external field lines caused by this motion
amplifies the current along these field lines giving rise to enhanced
magnetospheric scattering of X-rays from below, and in the case of open field
lines, increased torque on the star.  Within the context of this model, the
delayed spectral response would indicate that the open field lines near the
magnetic poles were first affected by the twist causing the sudden increase in
torque.  If only a small bundle of field lines were initially involved, the
phase-averaged energy spectrum would not be significantly altered.  Assuming
the current along closed field lines gradually increased over the next several
years, so would the isotropic scattering of X-rays, and, consequently, the
spectral hardness.

\begin{figure}[!h]

\centerline{
\psfig{file=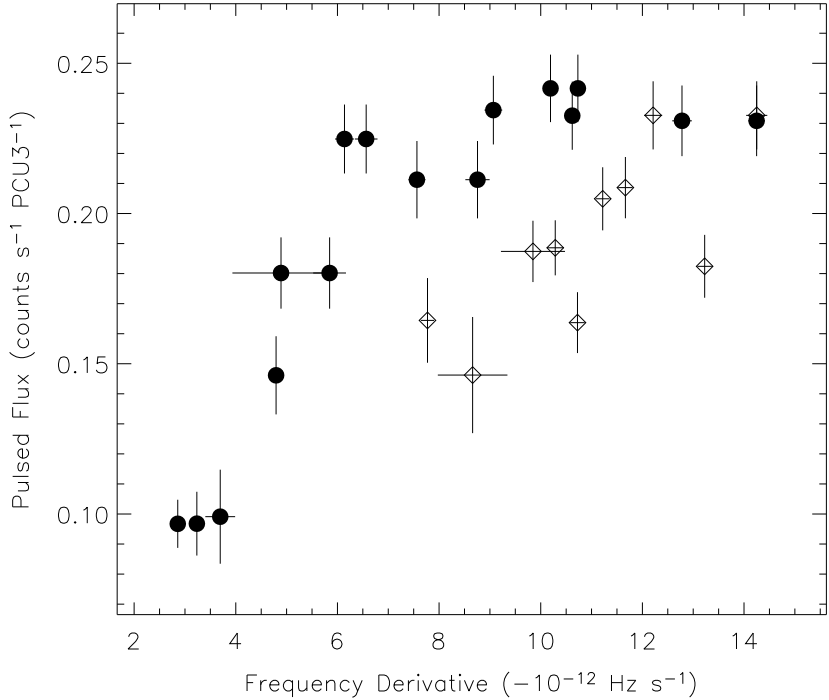,height=4.0in}}

\caption{Frequency derivative versus pulsed flux for \sgra\ during the epoch
surrounding the burst active episode in 2004$-$2005 (MJD 53050$-$53610).  Open
diamonds indicate measurements preceding the peak in torque on MJD 53209. 
Filled circles indicate measurements following this peak.  See text for
details. \label{fig:flux_torque}}

\end{figure}

The pulsed flux of \sgra\ correlates with the torque in the months surrounding
the giant flare.  In Figure~\ref{fig:flux_torque}, we show the frequency
derivative (i.e.\ torque) versus pulsed flux between MJD 53050 and 53610.  The
Spearman rank order correlation coefficient for this data set is 0.66 which
would be expected assuming the null-hypothesis (i.e.\ no correlation) with a
probability of 6 $\times$ 10$^{-4}$.  The measurements leading up to the peak
in torque on MJD 53209 are indicated by open diamonds and the post-peak
measurements are indicated by filled circles.  There is some evidence that the
decline in torque is more rapid than the pulsed flux decline since the filled
circles reside systematically higher than the open diamonds over the same range
in frequency derivative.  Within the model of Thompson et al.\ (2000, 2002),
the current flowing along open field lines determines the spin-down rate of the
neutron star.  This correlation suggests that the pulsed emission of the SGR is
shaped by these currents in the outer magnetosphere.  We note that no such
correlation was seen in the epoch surrounding the sudden torque change in
2000.  It is not clear why this correlation exists only over certain time
intervals.



Another consequence of crustal twisting would be an increase in burst activity
since the bursts are believed to be triggered by crustal fractures and/or
magnetospheric reconnection events.  We find that the burst frequency reached a
maximum shortly before the peak in spin-down rate in 2004.  This delay of
$\sim$2 months could reflect the time scale at which the twist propogates from
the stellar surface, presumably where the bursts originate, out to the light
cylinder where the torque on the star is influenced.  Curiously, there is no
sudden rise in burst activity at the time of the torque change in 2000.  

Magnetospheric currents strongly influence the pulsed intensity and morphology
of the persistent X-ray emission from magnetars (Thompson et al.\ 2002). 
Similar to the giant flare from \sgrc, the December 27$^{\rm th}$ flare from
\sgra\ had a lasting impact upon the pulse morphology of the X-ray emission
which has persisted for several months following the flare.  The fact that the
pulse profile of \sgra\ became more complex following the flare while the
\sgrc\ pulse profile simplified suggests that these current distributions can
become more complex or simplify as a consequence of the flare.  This
observation in combination with the sustained phase-averaged flux following
both flares supports the assertion made from flare energetics (e.g.\ Hurley et
al.\ 2005) that the ultimate energy source for these flares is likely internal
to the star as opposed to relaxation of external currents.

The spin frequency evolution of \sgra\ leading up to and following the giant
flare was significantly different than the spin behavior seen in \sgrc\ at the
time of its flare.  There was good circumstantial evidence to suggest that
\sgrc\ underwent a sudden change in frequency of $\frac{\Delta\nu}{\nu} \approx
-1 \times 10^{-4}$ at the time of the 1998 August 27 giant flare (Woods et al.\
2001).  Extrapolation of the pre-flare and post-flare spin ephemerides showed a
clear mismatch in frequency at the time of the flare (Woods et al.\ 1999) and
the pulse phase during the flare did not agree with a backward extrapolation of
the post-flare pulse ephemeris (Palmer 2002).  However, this conclusion was not
definitive due to uncertainty in the pre-flare pulse frequency of \sgrc\ and
energy dependence of the pulse profile during the flare.  Assuming the
frequency change was genuine, Thompson et al.\ (2000) hypothesized that the
drop in spin frequency of at the time of the giant flare could have been caused
by a particle outflow during the flare.  In fact, the transient radio nebula
detected in \sgrc\ (Frail, Kulkarni \& Bloom 1999) supports the assertion of a
particle outflow.  Assuming the outflow is restricted to a  fraction $f$ of the
surface, and accounting for the dependence of the outflows angular momentum on
the latitude $\theta$, we find

\begin{equation}
{\Delta \nu \over \nu} \simeq 1 \times 10^{-4} \, 
\left({\Delta E\over 10^{44}~{\rm erg}}\right)^{1/2}
\,\left({\Delta t\over 100~{\rm s}}\right)^{1/2}\,
\,f^{1/2}\,\sin^2\theta\,
\left({B_\star\over 10~B_{\rm QED}}\right)
\,\left({V\over 0.2 c}\right)^{-3/2} .
\end{equation}

\noindent where $\Delta E$ is the total kinetic energy of the particles blown
off the surface during the flare, $\Delta t$ is the time scale of the outflow,
$B_{\star}$ is the surface field strength, and $V$ is the velocity of the
outflowing particles.  Thompson et al.\ approximated the particle outflow
parameters and assumed isotropic emission to match the inferred frequency
change.

The \sgra\ flare was $\sim$2 orders of magnitude more luminous in
$\gamma$-ray/X-ray emission (Palmer et al.\ 2005) and radio emission (Gaensler
et al.\ 2005).  Moreover, extensive observations of the radio nebula provided
measurements of the ejected mass ($M \simeq 5 \times 10^{24}$ g [Gelfand et
al.\ 2005]), the total particle energy ($\Delta E$ $\simeq$ 3 $\times$
$10^{44}$ ergs [Gelfand et al.\ 2005]), and the initial outflow velocity ($V
\simeq 0.7c$ [Taylor et al.\ 2005]).  For a surface dipole magnetic field
strength of $\sim$10$^{15}$ G inferred from the pulse timing parameters, the
expected $\frac{\Delta \nu}{\nu}$ for \sgra\ was 

\begin{equation}
{\Delta \nu \over \nu} \simeq 5 \times 10^{-5} \, 
\left({\Delta t\over 100~{\rm s}}\right)^{1/2}\,
\,f^{1/2}\,\sin^2\theta.
\end{equation}

\noindent An ejection of this much mass from the surface of the star without
producing a change in $\frac{\Delta \nu}{\nu} > 5 \times 10^{-6}$ would require
that the particle outflow proceeded rapidly ($\sim$1 s) and/or the mass was
expelled along the spin axis of the star.  The non-spherical outflow of the
material (Taylor et al.\ 2005; Fender et al.\ 2006; Granot et al.\ 2006)
suggests that one or both of these requirements were met.

One of the most obvious differences between the effects of the giant flares of
\sgra\ and \sgrc\ on the persistent X-ray emission is the relative timing of
the observed changes.  In \sgrc, the giant flare preceded and almost certainly
triggered a flux enhancement and pulse profile change.  For \sgra, the
spectrum, flux and pulse profile were already changing several months before
the flare, although the most significant pulse morphology transition occurred
during/after the flare.  In \sgrc, there was no detected change in torque
preceding the flare whereas \sgra\ showed dramatic changes months before the
flare epoch.  The one pre-flare phenomenon common to both SGRs was the onset of
intense burst activity in the months preceding their respective flares.  It
appears that the presence of intense burst activity is a necessary, but not
sufficient condition to predict giant flares.  There are some counter-examples
where intense burst activity did not result in a giant flare (e.g.\ \sgra\ in
1984 and 1996), although there could be a burst-intensity threshold that must
be met in order to trigger a flare.  The burst rate from \sgra\ in the months
preceding the giant flare was higher and more energy was released than in
previously recorded outbursts.  With only two well-studied examples, it is
difficult to draw any firm conclusions on magnetar behavior that is predictive
of a giant flare.  However, since all persistent emission parameters behaved
differently in the years leading up to the flares produced by these two SGRs,
it appears that the burst activity is the most promising metric to use.

\acknowledgments{\noindent {\it Acknowledgements} --  }  The authors thank the
anonymous referee for carefully reading the manuscript and providing useful
comments that improved the quality of the paper.  We also thank Chris Thompson,
Rob Duncan, Sandro Mereghetti and Andrea Tiengo for useful comments. We thank
the \rxte\ planners Evan Smith and Divya Pierea for their careful planning of
the extensive \rxte\ observations.  PMW is grateful for support from SAO
through grant GO4-5073X and NASA through grant NNG~04GO50G. MHF acknowledges
support through grant NNG~04GO50G.  CK, SKP, and EG are grateful for support
from NASA through the Long Term Space Astrophysics (LTSA) grant NAG~5-9350.  KH
is grateful for support from the NASA LTSA program, grants FDNAG5-11451 and
NAG5-13080.

\newpage

\begin{center}
{\bf Appendix: Pulse Cycle Count Fitting Technique}
\end{center}

We present here a method for estimating the pulse cycle counts between pulse phase
measurements, and for quanitifing the uniqueness of the cycle counts determined. 
This method is based on $\chi^2$ fits of a set of pulse phase and frequency measurements
using a polynomial phase model. An upper threshold value for the fit statistic
$\chi^2_{max}$ is choosen, and all combinations of integer cycle counts which
result in minimum $\chi^2$ values below this threshold are found using a tree search. 
If the lowest value of minimum $\chi^2$ is sufficiently separated from the next-lowest value, then the
the cycle counts associated with lowest value $\chi^2$ are uniquely determined. Otherwise
the cycle counts are ambiguous, with several combinations of cycle counts being possible. 
 
For a short enough span of measurements (assuming no glitch has occured) we can model 
pulse phases $\phi_i$ and frequencies $\nu_i$ measured at barycentric times $t_i$ as 
\begin{eqnarray}
   \phi_i &=& \sum_{j=0}^{M} a_j(t_i-T_0)^j+\sum_{k=0}^{N-1} h(t_i-\tau_k) \nonumber\\
   \nu_i &=& \sum_{j=1}^{M} a_j j(t_i-T_0)^{j-1} \label{a1}
\end{eqnarray}
were the $a_j$ are the coeffients of an $M^{\rm th}$ order polynomial in time, $T_0$ is the 
time origin of the polynomial, the $n_k$ are integer offsets correcting any cycle slips,
which are applied at times $\tau_k$ which are taken to be between measurements,
and $h(x)$ is the Heavyside function which is 0 for $x<0$ and 1 for $x>0$. 

The $\chi^2$ of the fit with this model may be written as
\begin{equation}
\chi^2 = \left| 
    \left( \begin{array}{ccc}
         A          & H & {\bf z} \\
         \acute{A}  & 0 & {\bf \acute{z}} 
    \end{array} \right)
    \left( \begin{array}{r}
        {\bf a} \\ {\bf n} \\ -1 
    \end{array} \right)
    \right|^2 \label{a2}
\end{equation}
where $A$, $\acute{A}$ and $H$ are arrays with elements
\begin{equation}
A_{ij}  = (t_i-T_0)^j/\sigma_{\phi_i},  
~~H_{ik} = h(t_i-\tau_k)/\sigma_{\phi_i},
~~\acute{A}_{ij} =  j(t_i-T_0)^{j-1}/\sigma_{\nu_i},
\label{a3}\end{equation}
${\bf z}$ and ${\bf \acute{z}}$ are vectors with elements
\begin{equation}
z_i = \phi_i/\sigma_{\phi_i}, ~~\acute{z}_i = \nu_i/\sigma_{\nu_i},\label{a4}
\end{equation}
${\bf a}$ is the vector of polynomial coefficients, and ${\bf n}$ is the vector of integer
cycle slip offsets, with  $\sigma_{\phi_i}$ and $\sigma_{\nu_i}$ being the phase and
frequency errors. 

We will assume that for fixed cycle counts this fit is overdetermined. 
Using Householder transformations (a sequence of reflections of the column vectors) we can 
transform the matrix in equation \ref{a2} into upper-triangular form. 
The $\chi^2$ of the fit then has the form
\begin{eqnarray}
\chi^2 &=& \left| 
    \left( \begin{array}{ccc}
         Q & R & {\bf s} \\
         0  & U & {\bf v} \\
         0 & 0 & w 
    \end{array} \right)
    \left( \begin{array}{r}
        {\bf a} \\ {\bf n} \\ -1 
    \end{array} \right)
    \right|^2 \nonumber \\
  &=& | Q {\bf a}+R{\bf n}-{\bf s}|^2
      +|U{\bf n}-{\bf v}|^2+w^2~~.\label{a5}
\end{eqnarray}
For a given set of cycle counts ${\bf n}$ the estimates for the polynomal coefficents are
\begin{equation}
\hat{\bf a} = Q^{-1}({\bf s}-R{\bf n})~~{\rm with}~~
   {\rm covar}(\hat{\bf a},\hat{\bf a}) = (Q^TQ)^{-1} \label{a6}
\end{equation}
and the minimum $\chi^2$ is
\begin{equation}
   \chi^2({\bf n}) =  |U{\bf n}-{\bf v}|^2+w^2~~.\label{a7}
\end{equation}
Our strategy for searching for values of $\chi^2({\bf n})$ less than $\chi^2_{max}$ 
relies on the fact that $U$ is by construction an upper triangular square matrix. We
may therefore write
\begin{equation}
   \chi^2({\bf n}) = \sum_{j=0}^{N-1} \left(
          \sum_{k=j}^{N-1} U_{jk} n_k-w_j\right)^2+w^2 \label{a8}
\end{equation}
Let us suppose we have a list of all sets of offsets $n_{i+1} \ldots n_{N-1}$
that satisfy
\begin{equation}
       \chi^2_{i+1} \equiv \sum_{j=i+1}^{N-1} \left(
            \sum_{k=j}^{N-1} U_{jk} n_k-w_j\right)^2+w^2 < \chi^2_{max}~~.
    \label{a9}
\end{equation}
Then for one of these sets the choices of $n_i$ for which
$\chi^2_i < \chi^2_{max}$ is given by
\begin{equation}
\left|U_{ii}n_i-\left(w_i-\sum_{k=i+1}^{N-1}U_{ik}n_k\right)\right|
< \sqrt{\chi^2_{max}-\chi^2_{i+1}} ~~. \label{a10} 
\end{equation}
Thus we may start by finding $n_{N-1}$ that satisify equation \ref{a10} for $i=N-1$ (with
$\chi^2_N = w^2$), and work toward $i=0$ building a tree of cycle count solutions,
with twigs dying off when equation \ref{a10} has no integer solutions, and branching
when there are multiple solutions.

This search can be made much more efficient by introducing a transformed set of cycle
count offsets $\acute{{\bf n}}$. From equation \ref{a10} it is clear that number of 
solutions will increase dramatically if for some $i$, $|U_{ii}|$ is small. If we 
where estimating continuous rather then integer variables $n_i$, then $|U_{ii}|^{-1}$ 
would be the
standard deviation on $n_i$ for $n_{i+1},\ldots,n_{N-1}$ fixed. This can be large
either because the volume of the solution space is large, or because $n_i$ is
highly correlated with some other offsets in $n_0,\ldots,n_{i-1}$. If the latter is
the case, then there may be a large number of intermediate cycle count sets,
but only one at the end of the search. To reduce such correlations we
construct the covariance matrix for the cycle count offsets $P = (U^TU)^{-1}$
and look for linear transformations $T$, with
\begin{eqnarray}
  \acute{{\bf n}} &=& T{\bf n} \nonumber \\
  \acute{P} &=& TPT^T \label{a12}
\end{eqnarray}
that reduce the size of the diagonal elements of $\acute{P}$ relative to $P$. T is required to
have integer elements, and an inverse with integer elements, so that all possible
sets of integer offsets ${\bf n}$ can be represented by integer $\acute{{\bf n}}$.
We construct $T$ from a series of elementary transformations. First we look for
pairs $i$ and $j$ such that the transformation $\acute{n}_i = n_i+m*n_j$ for
some integer $m$ (with no other offsets changed) results in $\acute{P}_{ii} < P_{ii}$.
We continue transforming the offset vector ${\bf n}$ until no other transformations
of this type are possible.  Next transformations of the form 
$\acute{n}_i = {\bf k}\cdot{\bf n}$, where $k_i=1$ and the other values are integer, are
looked for. First the optimal real values of $k_j$ for $j \neq i$ are computed, and then these
are set to the nearest integer. If this transformation results in $\acute{P}_{ii} < P_{ii}$,
then it is applied. Transformations of this form are search for until no more can be found.
Then the cycle offsets are permuted so that $\acute{P}_{ii}$ decreases from largest 
to smallest with increasing $i$, so that the largest increases $\chi^2$, and the least
branching occurs at the beginning of the search. To incorporate these transformed 
cycle counts, the array $H$ in equation \ref{a2} needs to be replaced by $HT^{-1}$ prior to
the triangularization which results in equation \ref{a3}.

\end{document}